\documentclass[11pt]{article}

\usepackage{jheppub}

\usepackage[capitalise]{cleveref}
\usepackage{hyperref}
\usepackage{amsfonts,amsmath,amsthm,amssymb,amsbsy}
\usepackage{mathtools}
\usepackage{bm,bbm}
\usepackage{extarrows}
\newcommand{\overbar}[1]{{\mkern 1.5mu\overline{\mkern-1.5mu#1\mkern-1.5mu}\mkern 1.5mu}}
\usepackage{import}
\usepackage{xifthen}
\usepackage{pdfpages}
\usepackage{transparent}

\usepackage{fancyhdr}
\newtheorem{theorem}{Theorem}

\crefname{conjecture}{Conjecture}{Conjectures}

\newtheoremstyle{colon}
{}
{}
{}
{}
{\bfseries}
{:}
{ }
{}
\theoremstyle{colon}
\newtheorem{example}{Example}

\usepackage{tcolorbox}

\usepackage{enumitem}

\title{Pushforwards via Scattering Equations with Applications to Positive Geometries}

\author{Tomasz \L ukowski,}\emailAdd{t.lukowski@herts.ac.uk}
\author{Robert Moerman,}\emailAdd{r.moerman@herts.ac.uk}
\author{and Jonah Stalknecht}\emailAdd{j.stalknecht@herts.ac.uk}

\affiliation{Department of Physics, Astronomy and Mathematics, \\ University of Hertfordshire, \\  Hatfield, Hertfordshire, AL10 9AB, United Kingdom}

\abstract{In this paper we explore and expand the connection between two modern descriptions of scattering amplitudes, the CHY formalism and the framework of positive geometries, facilitated by the scattering equations. For theories in the CHY family whose $S$-matrix is captured by some positive geometry in the kinematic space, the corresponding canonical form can be obtained as the pushforward via the scattering equations of the canonical form of a positive geometry defined in the CHY moduli space. In order to compute these canonical forms in kinematic spaces, we study the general problem of pushing forward arbitrary rational differential forms via the scattering equations. We develop three methods which achieve this without ever needing to explicitly solve any scattering equations. Our results use techniques from computational algebraic geometry, including companion matrices and the global duality of residues, and they extend the application of similar results for rational functions to rational differential forms.}

\begin{document}

\maketitle

\section{Introduction}

Over the past decade, the scattering equations have proven to be fundamental to the study of scattering amplitudes. They find their origins in twistor string theory where novel formulae for the complete tree-level $S$-matrix of $\mathcal{N}=4$ supersymmetric Yang-Mills (SYM) theory \cite{Witten:2003nn,Roiban:2004vt,Roiban:2004yf} and $\mathcal{N}=8$ supersymmetric gravity \cite{Cachazo:2012da,Cachazo:2012kg,Cachazo:2012pz} were derived. These formulae consist of integrals over the moduli space of curves from the $n$-punctured Riemann sphere to the relevant kinematic space (either twistor space or momentum space) and were later subsumed by the Cachazo--He--Yuan (CHY) formalism \cite{Cachazo:2013gna,Cachazo:2013hca,Cachazo:2013iaa}. In this latter setting, further simplification comes from the fact that amplitudes are expressed as integrals directly over the moduli space of the $n$-punctured Riemann sphere. These moduli space integrals fully localize to the solutions of a set of algebraic equations, called the scattering equations, which connect points in the kinematic space to points in the moduli space. Moreover, the scattering equations are relevant to the scattering of massless particles in arbitrary dimensions \cite{Cachazo:2013hca} and their application extends to a broad range of quantum field theories \cite{Cachazo:2014nsa,Cachazo:2014xea}. Theories which share the same kinematic space have the same scattering equations, making them relatively universal, and the resulting scattering amplitudes differ due to the different integrands in the CHY formulae. 

There is however an important caveat when studying the scattering equations: since they are coupled algebraic equations, they are difficult to solve and the explicit form of their solutions are known only for very simple examples. This poses a non-trivial obstacle for algebraically evaluating CHY integral formulae, since such an evaluation amounts to summing the CHY integrand multiplied by some Jacobian factor over all solutions to the scattering equations. Nevertheless, it is possible to circumvent this problem by using the powerful machinery of computational algebraic geometry to evaluate amplitudes without ever needing to explicitly solve the scattering equations. This relies on the theory of Gr\"obner bases which has already been successfully applied in the context of scattering equations in \cite{Huang:2015yka, Cardona:2015eba,Cardona:2015ouc,Sogaard:2015dba}. In particular, two methods have been explored: one involving companion matrices \cite{Huang:2015yka} and the other involving the global duality of residues \cite{Sogaard:2015dba}.

\begin{figure}
		\centering		
		\includegraphics[width=0.6\textwidth]{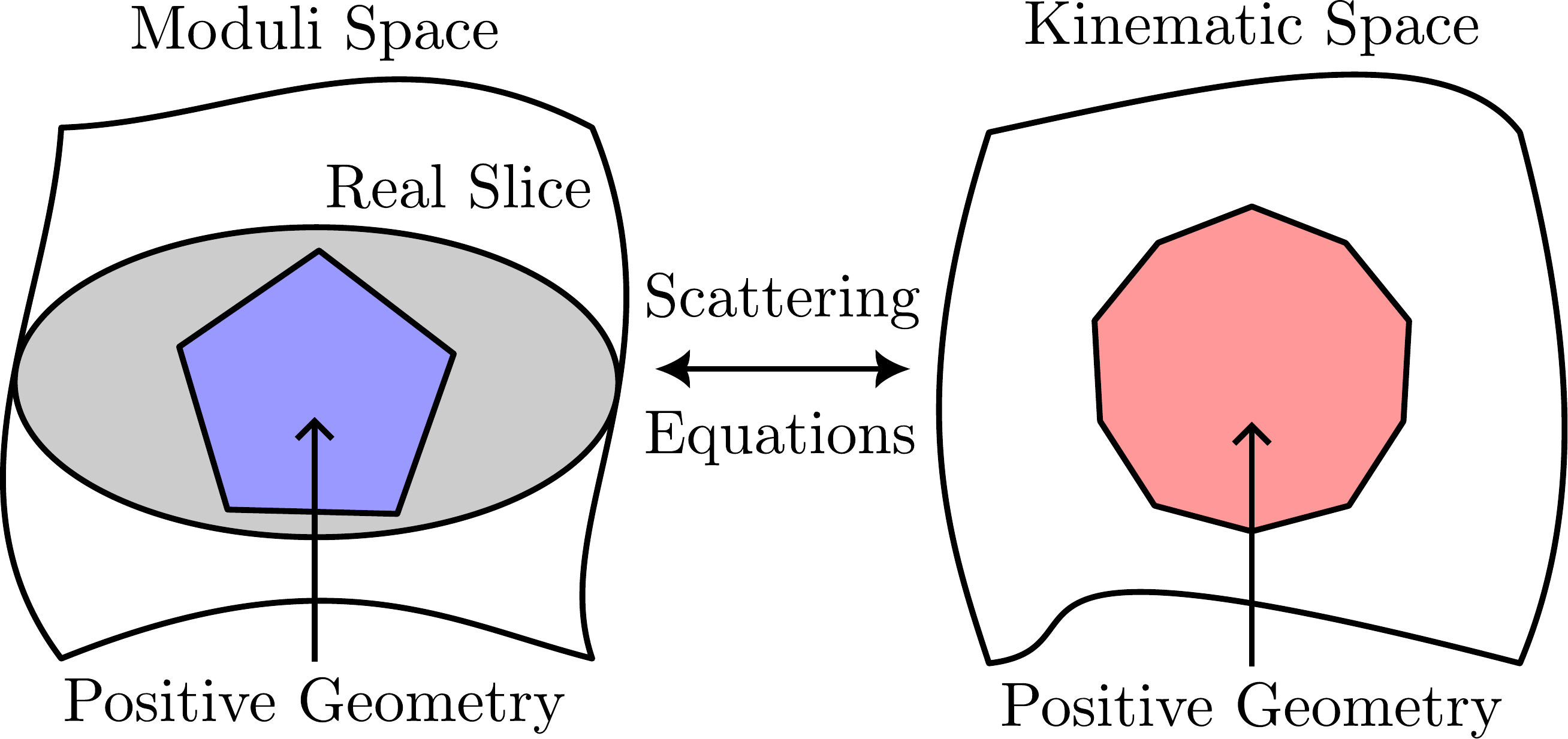}
		\caption{The scattering equations connect the CHY formalism to the framework of positive geometries. The canonical form of the positive geometry in the kinematic space is given by the pushforward through the scattering equations of the canonical form of the positive geometry in the moduli space.}
		\label{fig:connection}
\end{figure}

In a parallel development, scattering amplitudes for a subset of theories in the CHY family have also been described in the framework of positive geometries \cite{Arkani-Hamed2014,Arkani-Hamed:2017mur,Damgaard:2019ztj,He2022,Huang2021}. In this positive geometry approach, amplitudes are encoded by a canonical rational differential form associated to some positive region in the kinematic space. This positive region can be understood as the image of a positive region in the real slice of the CHY moduli space through the scattering equations \cite{Arkani-Hamed:2017mur,He2022}; the latter positive region also defines a positive geometry. Remarkably, the canonical form in the kinematic space can then be obtained as the pushforward through the scattering equations of the canonical form in the CHY moduli space \cite{Arkani-Hamed:2017mur,He2022}. This connection between the CHY formalism and the framework of positive geometries, mediated by the scattering equations, is depicted in \cref{fig:connection}.

It is therefore natural to ask if the commutative algebra methods mentioned above can be used to compute the pushforward of rational differential forms since the pushforward is also defined by summing over solutions to the scattering equations. In this paper we answer the question in the affirmative and explore methods for calculating pushforwards based on computational algebraic geometry. We derive expressions for the rational function prefactors of pushforwards which are directly amenable to the two previously named methods which have been developed for summing rational functions over solutions to the scattering equations. For the method that exploits the global duality of residues, we show that important simplifications occur when calculating the pushforward of top-dimensional rational differential forms. Moreover, we give a new formula for computing pushforwards as a sum over permutations which uses companion matrices and their derivatives with respect to kinematic parameters, and we provide an efficient algorithm for computing these derivatives numerically.

The paper is organised as follows. We begin by establishing notation and recalling a few facts from commutative algebra which will be relevant for this paper in \cref{sec:preliminaries}. \cref{sec:comp-matrices,sec:alternative-companion-matrices} focus on the method of companion matrices. More specifically, in \cref{sec:comp-matrices} we generalize the method used for summing functions over solutions to the scattering equations to the case of differential forms. In \cref{sec:alternative-companion-matrices} we present a new formula for calculating pushforwards using derivatives of companion matrices and we provide an efficient algorithm for finding these derivatives numerically. Thereafter, in \cref{sec:global-residue} we discuss how pushforwards can be calculated by leveraging the global duality of residues and we extend the methods previously developed for functions. We discuss the computational efficiency of our methods in \cref{sec:computation} and we consider some non-trivial applications of our methods to tree-level amplitudes in \cref{sec:examples}. We end the paper with \nameref{sec:conclusions} and we collect some important theorems in \cref{sec:theorems}.

\section{Mathematical Preliminaries}\label{sec:preliminaries}

\paragraph{Ideals and Varieties.}
Throughout this paper we use the language of commutative algebra to describe the scattering equations. We focus on the theory of ideals in polynomial rings with rational function coefficients, and use the standard theory of Gr\"obner basis and its applications to elimination theory. In the following we will assume that the reader is familiar with basic notions related to Gr\"obner bases which can be found e.g.\ in \cite{cox2013ideals}.

To start with, let $\mathbb{C}(\bm{a})[\bm{z}]\coloneqq\mathbb{C}(a_1,\ldots,a_m)[z_1,\ldots,z_n]$
be the ring of polynomials in $\bm{z}=(z_1,\ldots,z_n)$ whose coefficients are rational functions in $\bm{a}=(a_1,\ldots,a_m)$ with coefficients in $\mathbb{C}$, and suppose $n\le m$. We will refer to $\bm{z}$ as $z$-variables and $\bm{a}$ as $a$-variables. Since any polynomial $f\in \mathbb{C}(\bm{a})[\bm{z}]$ is a function of $z$-variables as well as a function of $a$-variables we will sometimes write $f(\bm{z};\bm{a})$ to emphasise this dual dependence while at other times we will simply write $f(\bm{z})$. Moreover, given a point $\bm{a}\in\mathbb{C}^m$ at which all of the coefficients of $f$ are analytic, we will use the notation $f(\bullet;\bm{a})$ to denote the polynomial in $\mathbb{C}[\bm{z}]\coloneqq\mathbb{C}[z_1,\ldots,z_n]$ defined by $f(\bullet;\bm{a}):\mathbb{C}^n\to\mathbb{C},\bm{z}\mapsto f(\bm{z};\bm{a})$.

In the context of the scattering equations, the $z$-variables are interpreted as coordinates in a chart in $\mathbb{C}^n$ of some moduli space (i.e.\ the positions of punctures on the Riemann sphere) and the $a$-variables as coordinates in a chart in $\mathbb{C}^m$ of some kinematic space (e.g.\ the space of Mandelstam variables for $n$ particles). The scattering equations are then a system of equations which implicitly define a map from the moduli space to the kinematic space.
Importantly, the scattering equations can always be expressed in polynomial form as a set of polynomials $f_1,\ldots,f_n\in\mathbb{C}(\bm{a})[\bm{z}]$ and therefore they define an \emph{ideal}
\begin{align}\label{eq:pf-ideal}
	\mathcal{I}\coloneqq\left\langle f_1,\ldots,f_n\right\rangle\subseteq\mathbb{C}(\bm{a})[\bm{z}]\,.
\end{align}
In order to elucidate the properties of this ideal, let us distinguish a few special subsets of $a$-variables. Firstly, let $A\subseteq\mathbb{C}^m$ denote the subset of $a$-variables for which all the coefficients of $f_1,\ldots,f_n$ are analytic. For each $\bm{a}\in A$, we define the ideal
\begin{align}\label{eq:pf-ideal-a}
	\mathcal{I}(\bm{a})\coloneqq\left\langle f_1(\bullet;\bm{a}),\ldots,f_n(\bullet;\bm{a})\right\rangle\subseteq\mathbb{C}[\bm{z}]\,.
\end{align}
We further define the following two subsets of $A$.
\begin{itemize}
	\item We define $A_\text{zero}$ to be the subset of $A$ for which $\mathcal{I}(\bm{a})$ is a \emph{zero-dimensional ideal}, i.e.\ the \emph{complex affine variety}
	\begin{align}\label{eq:pf-variety-a}
	\mathcal{V}(\mathcal{I}(\bm{a}))\coloneqq\left\{\bm{\xi}=(\xi_1,\ldots,\xi_n)\in\mathbb{C}^n\,|\,\forall_{f\in \mathcal{I}}:f(\bm{\xi})=0\right\}\subseteq\mathbb{C}^n\,,
	\end{align}
	has finite cardinality.
	\item We define $A_\text{rad}$ to be the subset of $A$ for which $\mathcal{I}(\bm{a})$ is a \emph{radical ideal} by which we mean that $	\mathcal{I}(\bm{a}) = \sqrt{\mathcal{I}(\bm{a})}$ where	\begin{align}\label{eq:pf-ideal-a-radical}
		 \sqrt{\mathcal{I}(\bm{a})}\coloneqq\left\{r\in \mathbb{C}[\bm{z}]\,|\, \exists_{n\in\mathbb{Z}_{>0}}:r^n\in \mathcal{I}(\bm{a})\right\}\subseteq\mathbb{C}[\bm{z}]\,.
	\end{align}
\end{itemize}
We will assume that $A_\text{gen} \coloneqq A_\text{zero}\cap A_\text{rad}$ is non-empty; we will refer to points $\bm{a}\in A_\text{gen}$ as being \emph{generic} $a$-variables and we will refer to $\mathcal{I}$ as being a \emph{generically} zero-dimensional radical ideal. Indeed, it is well-known that the original Cachazo--He--Yuan (CHY) scattering equations have finitely many solutions \cite{Cachazo:2013iaa,Cachazo:2013gna,Dolan:2015iln}, and hence define a zero-dimensional ideal. Moreover, we will assume that the complex affine variety for $\mathcal{I}$
\begin{align}\label{eq:pf-variety}
	\mathcal{V}(\mathcal{I})\coloneqq\left\{\bm{\xi}=(\xi_1,\ldots\xi_n)\in\overbar{\mathbb{C}(\bm{a})}^n\,|\, \forall_{f\in \mathcal{I}}:f(\bm{\xi})=0\right\}\subseteq \overbar{\mathbb{C}(\bm{a})}^n\,,
\end{align}
is well-defined for generic $a$-variables and that it contains $d$ distinct points. We will enumerate these points by \smash{$\{\bm{\xi}^{(\alpha)}\}_{\alpha=1}^{d}$}. Here $\mathbb{C}(\bm{a})\coloneqq\mathbb{C}(a_1,\ldots,a_m)$ while \smash{$\overbar{\mathbb{C}(\bm{a})}$} denotes the algebraic closure of $\mathbb{C}(\bm{a})$. Importantly, the common zeros of $f_1,\ldots,f_n$ given by $\mathcal{V}(\mathcal{I})$ are functions of $a$-variables.

\paragraph{Notation.}
To streamline our presentation in the next sections, we collect the following definitions and notation (see \cite{cox2013ideals} for a more detailed introduction to Gr\"obner bases). We denote monomials in $z_1,\ldots,z_n$ as $\bm{z}^{\bm{\alpha}} = z_1^{\alpha_1}z_2^{\alpha_2}\cdots z_n^{\alpha_n}$. A \emph{monomial ordering} $\prec$ on $\mathbb{C}(\bm{a})[\bm{z}]$ is a total well-ordering on the monomials in $\bm{z}$ that is compatible with multiplication: if $\bm{z}^{\bm{\alpha}}\succ\bm{z}^{\bm{\beta}}$, then $\bm{z}^{\bm{\alpha+\gamma}}\succ\bm{z}^{\bm{\beta+\gamma}}$ for all $\smash{\bm{\gamma} \in \mathbb{Z}_{\ge0}^n}$. For a given polynomial in $\mathcal{I}$, its \emph{leading term} is the monomial which is largest with respect to $\succ$. We will give a brief introduction to three useful monomial orders. Throughout, we define $z_1\succ z_2\succ\ldots\succ z_n$.
\begin{itemize}
	\item \emph{Lexicographic (lex)} ordering: $\bm{z}^{\bm{\alpha}}\succ_{lex}\bm{z}^{\bm{\beta}}$ if the leftmost non-zero entry of $\bm{\alpha}-\bm{\beta}$ is positive.
	\item \emph{Graded lexicographic (grlex)} ordering: $\bm{z}^{\bm{\alpha}}\succ_{grlex}\bm{z}^{\bm{\beta}}$ if $\sum_{i=1}^n \alpha_i > \sum_{i=1}^n \beta_i$, and if $\sum_{i=1}^n \alpha_i = \sum_{i=1}^n \beta_i$, then $\bm{z}^{\bm{\alpha}}\succ_{grlex}\bm{z}^{\bm{\beta}}$ if the leftmost non-zero entry of $\bm{\alpha}-\bm{\beta}$ is positive.
	\item \emph{Graded reverse lexicographic (grevlex)} ordering: $\bm{z}^{\bm{\alpha}}\succ_{grevlex}\bm{z}^{\bm{\beta}}$ if $\sum_{i=1}^n \alpha_i > \sum_{i=1}^n \beta_i$, and if $\sum_{i=1}^n \alpha_i = \sum_{i=1}^n \beta_i$, then $\bm{z}^{\bm{\alpha}}\succ_{grevlex}\bm{z}^{\bm{\beta}}$ if the rightmost non-zero entry of $\bm{\alpha}-\bm{\beta}$ is negative.
\end{itemize}
In this paper, we will need to use several monomial orderings for different parts of the discussions. In general, the reader may assume that the choice of monomial ordering is irrelevant, unless we explicitly specify otherwise.

Given some monomial ordering $\prec$ on $\mathbb{C}(\bm{a})[\bm{z}]$, we denote by $\mathcal{G}_{\prec}(\mathcal{I})$ the \emph{unique reduced Gr\"{o}bner basis} of $\mathcal{I}$ with respect to $\prec$. A \emph{Gr\"obner basis} $\mathcal{G_\prec}(\mathcal{I})$ is defined as a collection of polynomials $\mathcal{G}_\prec(\mathcal{I})=\{g_1,\ldots,g_t\}\subseteq \mathcal{I}$ such that for every (non-zero) $f \in \mathcal{I}$ the leading term of $f$ is divisible by the leading term of some $g_i \in \mathcal{G}_\prec(\mathcal{I})$. Moreover $\mathcal{G}_\prec(\mathcal{I})$ generates the ideal: $\mathcal{I}=\langle g_1,\ldots,g_t \rangle$. For brevity, we will often write $\mathcal{G}$ instead of $\mathcal{G}_{\prec}(\mathcal{I})$. Since $\mathcal{I}$ is a zero-dimensional ideal, the \emph{quotient ring} $Q=\mathbb{C}(\bm{a})[\bm{z}]/\mathcal{I}$ is a finite-dimensional vector space over $\mathbb{C}(\bm{a})$ of dimension $\dim(Q) = |\mathcal{V}(\mathcal{I})|=d$. We denote by $\mathcal{B}_\prec(\mathcal{I})$, or simply $\mathcal{B}$, the set of all monomials $\bm{z}^{\bm{\alpha}} = z_1^{\alpha_{1}}\cdots z_n^{\alpha_{n}}\in \mathbb{C}(\bm{a})[\bm{z}]$ for $\bm{\alpha}=(\alpha_1,\ldots,\alpha_n)\in\mathbb{Z}^n_{\ge0}$ which are indivisible by any of the leading terms of polynomials in $\mathcal{G}$. The set $\mathcal{B}$ forms a basis for $Q$ because every polynomial in $\mathbb{C}(\bm{a})[\bm{z}]$ can be written as a $\mathbb{C}(\bm{a})$-linear combination of monomials in $\mathcal{B}$ modulo $\mathcal{I}$ via the division algorithm with respect to $\mathcal{G}$. We call $\mathcal{B}$ the \emph{standard (monomial) basis} for $Q$ and we label its elements by \smash{$\mathcal{B}=\{e_\alpha\}_{\alpha=1}^{d}$}. Given a polynomial $f\in \mathbb{C}(\bm{a})[\bm{z}]$, we denote the \emph{normal form} of $f$, i.e.\ the remainder of $f$ on division by $\mathcal{G}$, by $\overbar{f}^{\mathcal{G}}= \sum_{\alpha=1}^df_\alpha e_\alpha$ where the coefficients $f_\alpha\in\mathbb{C}(\bm{a})$ are rational functions in $a$-variables with coefficients in $\mathbb{C}$.

Throughout this paper, we will use $[n]$ to denote the set $\{1,2,\ldots,n\}$ and \smash{$\binom{[n]}{p}$} to denote the set of all $p$-element subsets of $[n]$. Importantly, if $I=\{ i_1,\ldots,i_p\}\in\binom{[n]}{p}$ then we always assume that $i_1<i_2<\ldots<i_p$.

\paragraph{Pullbacks.} In this paper we are interested in the operation of pushing forward differential forms. The pushforward is defined in terms of multiple pullback operations, so we will introduce the latter operation first.

Suppose we have a map $\bm{\phi}:\mathbb{C}^m\to\mathbb{C}^n;\bm{a}\mapsto \bm{z}=\bm{\phi}(\bm{a})=(\phi_1(\bm{a}),\ldots,\phi_n(\bm{a}))$ from $a$-variables to $z$-variables (i.e.\ from a chart of some kinematic space to a chart of some moduli space). Without loss of generality, let
\begin{align}\label{eq:omega}
	\omega = \underline{\omega}(\bm{z})\bigwedge_{i\in I}\mathrm{d}z_i=\underline{\omega}(\bm{z})\,\mathrm{d}z_{i_1}\wedge\cdots\wedge\mathrm{d}z_{i_p}\,,
\end{align}
be a $p$-form in $z$-variables (with $p\le n$) for some choice of indices $I=\{i_1,\ldots,i_p\}\in\binom{[n]}{p}$. Here $\underline{\omega}\in\mathbb{C}(\bm{z})$ and we refer to $\omega$ as a rational $p$-form. Then the \emph{pullback of $\omega$ through $\bm{\phi}$}, denoted by $\bm{\phi}^\ast\omega$, is defined as the evaluation of $\omega$ on $\bm{z}=\bm{\phi}(\bm{a})$:
\begin{align}
	\bm{\phi}^\ast\omega \coloneqq \omega\big|_{\bm{z}=\bm{\phi}(\bm{a})} = \underline{\omega}(\bm{\phi}(\bm{a}))\bigwedge_{i\in I}\mathrm{d}\phi_i(\bm{a}) = \underline{\omega}(\bm{\phi}(\bm{a}))\sum_{J\in\binom{[m]}{p}}\left|\frac{\partial\bm{\phi}}{\partial\bm{a}}\right|^I_J\bigwedge_{j\in J}\mathrm{d}a_j\,,
\end{align}
where \smash{${\partial{\bm{\phi}}}/{\partial\bm{a}}$} is the Jacobian matrix of partial derivatives \smash{${\partial\phi_i}/{\partial a_j}$} and \smash{$\big|\cdots\big|^{I}_{J}$} denotes the minor of $(\,\cdots)$ specified by the rows in $I$ and the columns in $J$. The pullback operation extends by linearity to arbitrary rational differential forms.

One can also define the pullback of $\omega$ through an ideal that implicitly defines such a map $\bm{\phi}$. To this end, let $\mathcal{I}\subseteq\mathbb{C}(\bm{a})[\bm{z}]$ be a generically zero-dimensional radical ideal and suppose that the corresponding complex affine variety $\mathcal{V}(\mathcal{I})$ generically contains a single point. Then the  \nameref{thm:shape} ensures that on the support of generic $a$-variables, the Gr\"{o}bner basis for $\mathcal{I}$ with respect to lex ordering takes a particularly convenient form:
\begin{align}\label{eq:pb-groebner-lex}
	\mathcal{G}_\text{lex}(\mathcal{I})=\left\{z_1-\phi_1(\bm{a}),\ldots,z_n-\phi_n(\bm{a})\right\},
\end{align}
where $\phi_1,\ldots,\phi_n\in\mathbb{C}(\bm{a})$ are analytic on $A_\text{gen}$ and they are constant functions with respect to $z$-variables. In this case
\begin{align}
	\mathcal{V}(\mathcal{I})=\{\bm{\phi}=(\phi_1,\dots,\phi_n)\in\mathbb{C}(\bm{a})^n\}\,,
\end{align}
and the \emph{pullback of $\omega$ through $\mathcal{I}$}, denoted by $\mathcal{I}^\ast\omega$, is given by
\begin{align}\label{eq:pb-form-definition}
	\mathcal{I}^\ast\omega \coloneqq\bm{\phi}^\ast\omega\,.
\end{align}

\begin{figure}[t]
	\centering
	\includegraphics[width=0.6\textwidth]{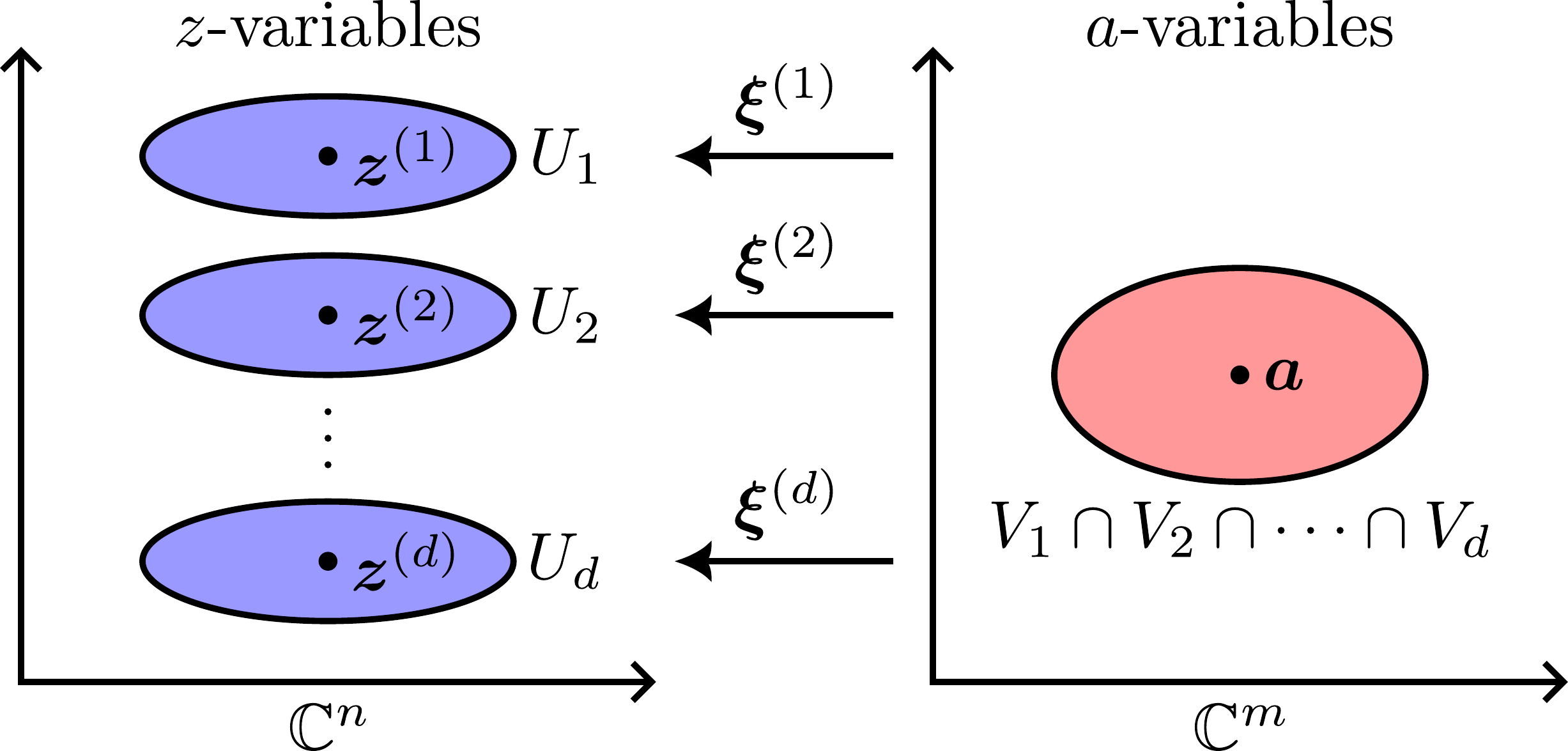}
	\caption{Around the point $\bm{a}\in\mathbb{C}^m$, the map $\bm{\psi}:\mathbb{C}^n\to\mathbb{C}^m$ has $d$ local inverse maps \smash{$\bm{\xi}^{(\alpha)}\coloneqq\bm{\psi}\big|_{U_\alpha}^{-1}:V_\alpha\to U_\alpha$} which take $\bm{a}$ to \smash{$\bm{z}^{(\alpha)}\coloneqq\bm{\xi}^{(\alpha)}(\bm{a})$}.}
	\label{fig:xi}
\end{figure}

\paragraph{Pushforwards.}
\begin{figure}[t]
	\centering
	\includegraphics[width=0.6\textwidth]{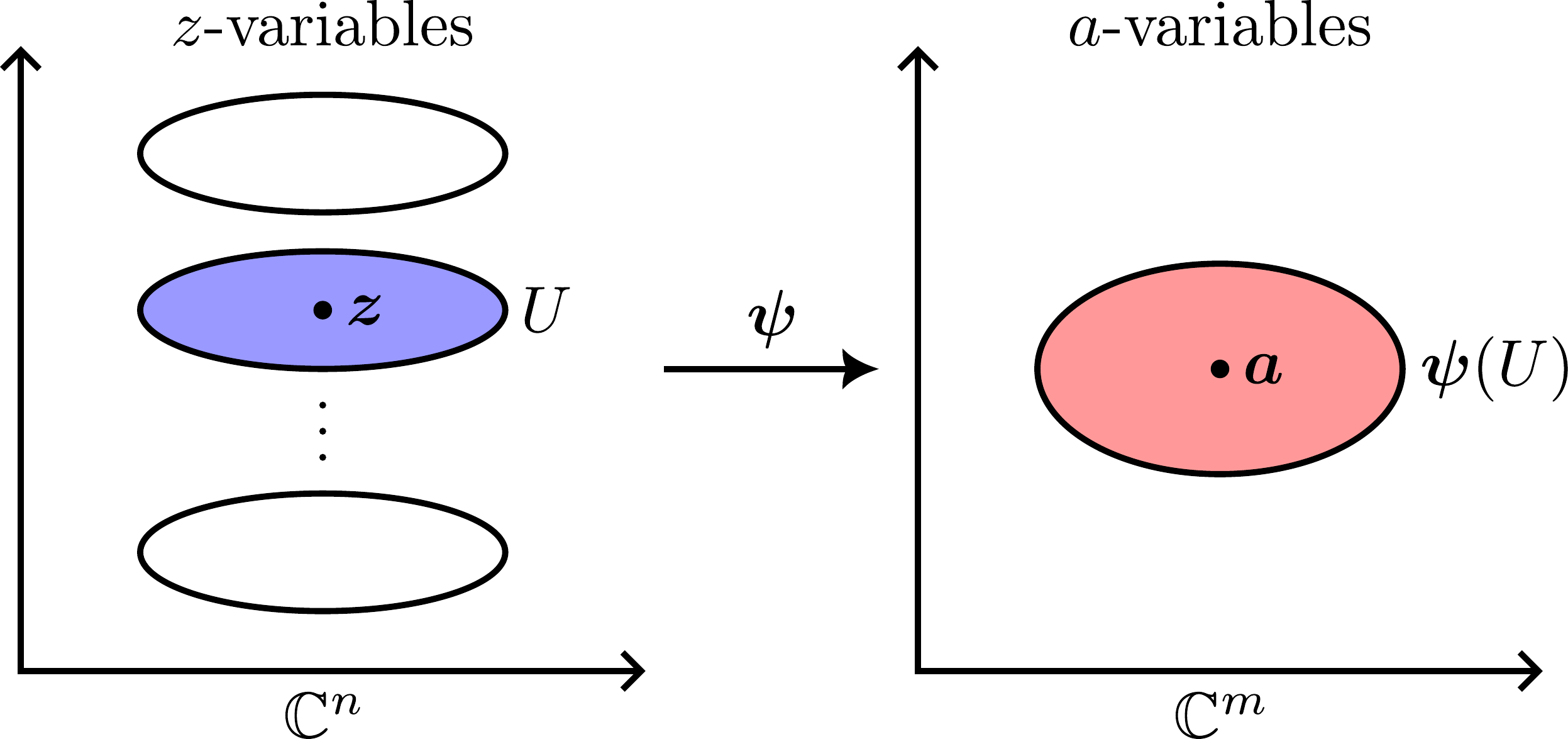}
	\caption{The map $\bm{\psi}:\mathbb{C}^n\to\mathbb{C}^m$ takes $\bm{z}$ to $\bm{a}=\bm{\psi}(\bm{z})$.}
	\label{fig:psi}
\end{figure}
We can also push our rational $p$-form $\omega$ given in \eqref{eq:omega} forward through a map defined in the direction opposite to $\bm{\phi}$. We define the pushforward as in \cite{Arkani-Hamed2017}, but we do not restrict ourselves to top-forms. Let $\bm{\psi}:\mathbb{C}^n\to\mathbb{C}^m;\bm{z}\mapsto\bm{a}=\bm{\psi}(\bm{z})$ be a map from $z$-variables to $a$-variables (i.e.\ from a chart of some moduli space to a chart of some kinematic space) as depicted in \cref{fig:psi}. We assume that $\bm{\psi}$ is a meromorphic map of degree $d$, by which we mean that (in general) a point $\bm{a}\in\mathbb{C}^m$ has $d$ preimages
\begin{align}
	\bm{\psi}^{-1}(\{\bm{a}\}) = \{\bm{z}^{(\alpha)}\}_{\alpha=1}^{d},
\end{align}
where \smash{$\bm{z}^{(\alpha)}\in\mathbb{C}^n$}. For each $\alpha=1,\dots,d$, there exists an open neighbourhood $U_\alpha\subseteq \mathbb{C}^n$ containing \smash{$\bm{z}^{(\alpha)}$} and an open neighbourhood $V_\alpha\subseteq\mathbb{C}^m$ containing $\bm{a}$ such that we can define the local inverse map $\bm{\xi}^{(\alpha)}\coloneqq\bm{\psi}\big|_{U_\alpha}^{-1}\colon V_\alpha \to U_\alpha$ where $\bm{z}^{(\alpha)}=\bm{\xi}^{(\alpha)}(\bm{a})$. These local inverse maps are depicted in \cref{fig:xi}. We then define the \emph{pushforward of $\omega$ through $\bm{\psi}$}, denoted by $\bm{\psi}_\ast\omega$ as the sum over all pullbacks of $\omega$ through \smash{$\bm{\xi}^{(\alpha)}$}:
\begin{align}
\bm{\psi}_\ast\omega\coloneqq\sum_{\alpha=1}^{d}\bm{\xi}^{(\alpha)\ast}\omega\,.
\end{align}
As was the case for the pullback, the definition of the pushforward extends by linearity to arbitrary rational differential forms.

\begin{figure}[t]
	\centering
	\includegraphics[width=0.6\textwidth]{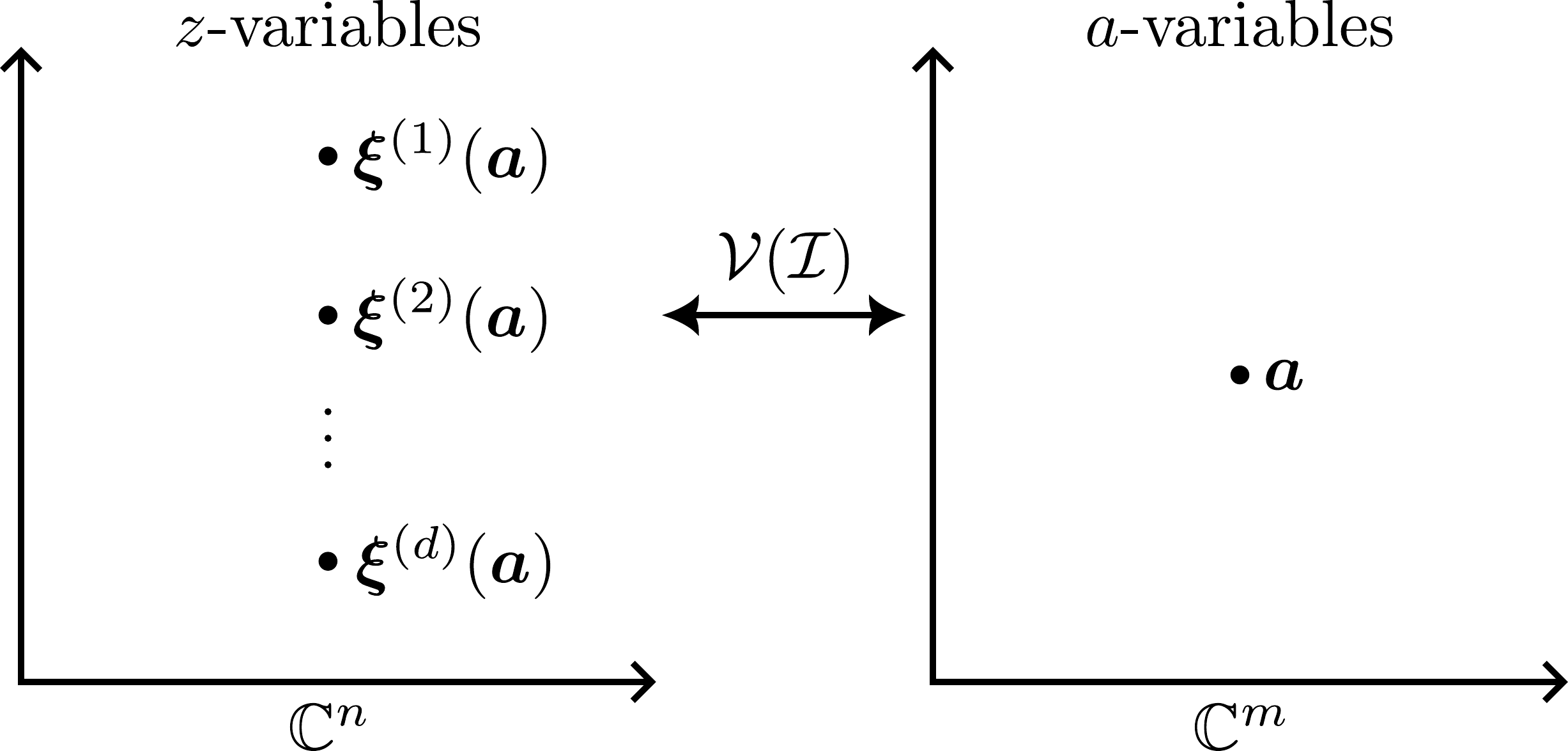}
	\caption{The variety $\mathcal{V}(\mathcal{I})=\{\bm{\xi}^{(\alpha)}\}_{\alpha=1}^d$ generically consists of $d$ maps in $\overbar{\mathbb{C}(\bm{a})}^n$ which take $\bm{a}$ to $\bm{\xi}^{(\alpha)}(\bm{a})$.}
	\label{fig:variety}
\end{figure}

In this paper, we are especially interested in computing the pushforward of rational differential forms through the common zeroes of a set of polynomial equations, i.e.\ through an ideal. Given a generically zero-dimensional radical ideal $\mathcal{I}=\langle f_1,\ldots,f_n\rangle \in\mathbb{C}(\bm{a})[\bm{z}]$, we define the pushforward of $\omega$ through $\mathcal{I}$, denoted by $\mathcal{I}_\ast\omega$, via
\begin{align}\label{eq:pf-form-definition}
\mathcal{I}_\ast\omega\coloneqq\sum_{\bm{\xi}\in\mathcal{V}(\mathcal{I})}\bm{\xi}^\ast\omega\,,
\end{align}
where the complex affine variety $\mathcal{V}(\mathcal{I})=\{\bm{\xi}^{(\alpha)}\}_{\alpha=1}^d$ generically consists of $d$ maps in $\overbar{\mathbb{C}(\bm{a})}^n$ as shown in \cref{fig:variety}. Comparing formulae \eqref{eq:pf-form-definition} and \eqref{eq:pb-form-definition}, we notice that a pullback through an ideal can be thought of as a special case of a pushforward through an ideal, where the relevant variety contains just one point.
To evaluate \eqref{eq:pf-form-definition} directly requires one to explicitly determine all points $\bm{\xi}\in\mathcal{V}(\mathcal{I})$ and, in general, this is impossible to do in closed form. Thus, the question we want to consider is this: \emph{can we compute the pushforward of $\omega$ through the common zeroes of $f_1,\ldots,f_n$ without explicitly determining these common zeros?}

There is a closely related problem which concerns summing a rational function $r\in\mathbb{C}(\bm{z})$, as opposed to a rational differential form, over all points $\bm{\xi}\in\mathcal{V}(\mathcal{I})$. Interpreting $r$ as a $0$-form, we see that this is equivalent to calculating the pushforward of $r$ through $\mathcal{I}$ as defined in \eqref{eq:pf-form-definition}. This simpler problem has been studied in the scattering amplitude literature, and there are two particular solutions known to the authors which (we will show) can be applied to rational differential forms. The first approach uses the machinery of \emph{companion matrices} \cite{Huang:2015yka,Cardona:2015eba,Cardona:2015ouc}, and the second uses the \emph{global duality of residues} \cite{Sogaard:2015dba}.

To address the problem at hand, we derive a formula for the rational function coefficients of the pushforward which is amenable to the above methods. Using the chain-rule \smash{$\mathrm{d}\xi_i = {\partial \xi_i}/{\partial a_j}\,\mathrm{d}a_j$}, one can rewrite \eqref{eq:pf-form-definition} as
\begin{align}\label{eq:pf-form-chain-rule}
	\mathcal{I}_\ast\omega=\sum_{J\in\binom{[m]}{p}}\left[\sum_{\bm{\xi}\in\mathcal{V}(\mathcal{I})}\underline{\omega}(\bm{\xi})\bigg|\frac{\partial\bm{\xi}}{\partial\bm{a}}\bigg|^{I}_{J}\right]\,\bigwedge_{j\in J}\mathrm{d}a_{j}\,,
\end{align}
where \smash{${\partial\bm{\xi}}/{\partial\bm{a}}$} is the Jacobian matrix of partial derivatives \smash{${\partial\xi_i}/{\partial a_j}\in\overbar{\mathbb{C}(\bm{a})}$}. The minor in \eqref{eq:pf-form-chain-rule} can be rewritten in a more convenient manner using the following observation. By definition, $f_\ell(\bm{\xi}(\bm{a});\bm{a})=0$ for each point $\bm{\xi}\in\mathcal{V}(\mathcal{I})$, each $\ell\in[n]$, and each $\bm{a}\in A_\text{gen}$. Consequently, for each $j\in[m]$
\begin{align}\label{eq:total-derivative-f-a}
0 = \frac{df_{\ell}}{da_j} = \frac{\partial f_\ell}{\partial a_j} + \frac{\partial f_\ell}{\partial z_i}\frac{\partial \xi_i}{\partial a_j} \implies \frac{\partial f_\ell}{\partial z_i}\frac{\partial \xi_i}{\partial a_j} = - \frac{\partial f_\ell}{\partial a_j}\,,
\end{align}
where function of $z$-variables are evaluated on $\bm{z}=\bm{\xi}(\bm{a})$. From \eqref{eq:total-derivative-f-a} it follows that
\begin{align}\label{eq:pf-jacobian-chain-rule}
\bigg|\frac{\partial\bm{\xi}}{\partial\bm{a}}\bigg|^{I}_{J}  = (-1)^p\, \bigg|\bigg[\frac{\partial\bm{f}}{\partial\bm{z}}(\bm{\xi})\bigg]^{-1}\frac{\partial\bm{f}}{\partial\bm{a}}(\bm{\xi})\bigg|^{I}_{J},
\end{align}
where \smash{${\partial\bm{f}}/{\partial\bm{z}}$} is the Jacobian matrix of partial derivatives \smash{${\partial f_\ell}/{\partial z_i}\in\mathbb{C}(\bm{a})[\bm{z}]$} and  \smash{${\partial\bm{f}}/{\partial\bm{a}}$} is the Jacobian matrix of partial derivatives \smash{${\partial f_\ell}/{\partial a_j}\in\mathbb{C}(\bm{a})[\bm{z}]$}. Whenever $\bm{a}\in A_\text{gen}$, the Jacobian matrix \smash{${\partial\bm{f}}/{\partial\bm{z}}\,(\bm{\xi}(\bm{a});\bm{a})$} is invertible, and therefore for the remainder of this paper we will restrict our attention to generic $a$-variables. Using \eqref{eq:pf-jacobian-chain-rule} we can rewrite \eqref{eq:pf-form-chain-rule} as
\begin{align}\label{eq:pf-form-with-coefficients}
	\mathcal{I}_\ast\omega=\sum_{J\in \binom{[m]}{p}}\mathcal{I}_\ast\underline{\omega}^{I}_{J}\,\bigwedge_{j\in J}\mathrm{d}a_{j}\,,
\end{align}
where
\begin{align}\label{eq:pf-coefficients-pushforward}
	\mathcal{I}_\ast\underline{\omega}^{I}_{J} \coloneqq \sum_{\bm{\xi}\in\mathcal{V}(\mathcal{I})}\underline{\omega}^{I}_{J}(\bm{\xi})\,,
\end{align}
and
\begin{align}\label{eq:pf-coefficients}
 \underline{\omega}^{I}_{J}(\bm{z};\bm{a})\coloneqq (-1)^p\,\underline{\omega}(\bm{z})	\bigg|\bigg[\frac{\partial\bm{f}}{\partial\bm{z}}\bigg]^{-1}\frac{\partial\bm{f}}{\partial\bm{a}}\bigg|^{I}_{J}\in\mathbb{C}(\bm{a})(\bm{z})\,.
\end{align}
Then, we have reduced the problem of calculating the pushforward of $\omega$ through $\mathcal{V}(\mathcal{I})$ to the problem of summing the rational functions \smash{$\underline{\omega}^{I}_{J}$} over all points $\bm{\xi}\in\mathcal{V}(\mathcal{I})$. In the next sections, we present three methods for computing \eqref{eq:pf-coefficients-pushforward} without needing to determine $\mathcal{V}(\mathcal{I})$. However, before proceeding to these methods, let us consider the following example in order to clarify some of the above discussions.
\begin{example}\label{ex:pf-formal}
	Consider the zero-dimensional ideal
	\begin{align}
		\mathcal{I}=\langle f_1,f_2\rangle=\langle a_1z_1+a_2z_2,a_3z_2^2-1 \rangle\in \mathbb{C}(a_1,a_2,a_3)[z_1,z_2]\,.
	\end{align}
	In this case, the corresponding complex affine variety is easily calculated to be
	\begin{align}
		\mathcal{V}(\mathcal{I})=\left\{\left( \frac{a_2}{a_1\sqrt{a_3}},-\frac{1}{\sqrt{a_3}}\right),\left( -\frac{a_2}{a_1\sqrt{a_3}},\frac{1}{\sqrt{a_3}}\right)\right\}.
	\end{align}
	Clearly $A_\text{zero} = \big\{(a_1,a_2,a_3)\in\mathbb{C}^3: a_1\ne 0 \text{ and } a_3\ne0\big\}$ which is precisely the set of $a$-variables for which the points in $\mathcal{V}(\mathcal{I})$ are well-defined and distinct.
	Suppose we wish to compute the push-forward of
	\begin{align}
		\omega=\mathrm{d}\log z_1\wedge \mathrm{d}\log z_1 = \underline{\omega}(\bm{z})\,\mathrm{d}z_1\wedge \mathrm{d}z_2\,,\qquad\text{where}\qquad\underline{\omega}(\bm{z})=\frac{1}{z_1z_2}\,.
	\end{align}
	Direct computation via \eqref{eq:pf-form-definition} yields
	\begin{align}\label{eq:ex-pf-form-definition}
		\mathcal{I}_\ast\omega = \frac{\mathrm{d}a_1\wedge \mathrm{d}a_3}{a_1a_3} - \frac{\mathrm{d}a_2\wedge \mathrm{d}a_3}{a_2a_3}\,.
	\end{align}
	This result can be also be obtained via \eqref{eq:pf-form-with-coefficients}. The Jacobian matrices are
	\begin{align}
		\frac{\partial\bm{f}}{\partial\bm{z}}=\begin{pmatrix}
			a_1 & a_2\\
			0 & 2 a_3z_2
		\end{pmatrix},\qquad \frac{\partial\bm{f}}{\partial\bm{a}}=\begin{pmatrix}
		z_1 & z_2 & 0\\
		0 & 0 & z_2^2
	\end{pmatrix}.
	\end{align}
	Notice that \smash{${\partial\bm{f}}/{\partial\bm{z}}$} is invertible only on $A_\text{zero}$.
	The coefficients \smash{$\underline{\omega}_{\{j_1,j_2\}}\coloneqq\underline{\omega}^{\{1,2\}}_{\{j_1,j_2\}}$}, calculated according to \eqref{eq:pf-coefficients}, are given by
	\begin{align}\label{eq:ex-pf-coefficients}
		\underline{\omega}_{\{1,2\}}=0\,,\qquad\underline{\omega}_{\{1,3\}}=\frac{1}{2a_1a_3}\,,\qquad \underline{\omega}_{\{2,3\}}=\frac{z_2}{2 a_1 a_3 z_1}\,.
	\end{align}
	Then summing \eqref{eq:ex-pf-coefficients} over all points $\bm{\xi}\in\mathcal{V}(\mathcal{I})$ yields
	\begin{align}\label{eq:ex-pf-coefficients-pushforward}
		\mathcal{I}_\ast\underline{\omega}_{\{1,2\}}=0\,,\qquad \mathcal{I}_\ast\underline{\omega}_{\{1,3\}} = \frac{1}{a_1a_3}\,,\qquad \mathcal{I}_\ast\underline{\omega}_{\{2,3\}} = -\frac{1}{a_2a_3}\,,
	\end{align}
	in agreement with \eqref{eq:ex-pf-form-definition}.
\end{example}
\noindent\cref{ex:pf-formal} highlights two interesting implications of \eqref{eq:pf-coefficients-pushforward}:
\begin{itemize}
	\item If \smash{$\big|{\partial\bm{f}}/{\partial\bm{a}}\big|{}^{I}_{J}=0$}, then \smash{$\underline{\omega}^{I}_{J}=0$} and hence \smash{$I_\ast\underline{\omega}^{I}_{J}=0$} for any rational differential form $\omega$. In this case, the vanishing of \smash{$\underline{\omega}^{I}_{J}$} and \smash{$\mathcal{I}_\ast\underline{\omega}^{I}_{J}$} is a consequence of the scattering equations themselves, and is independent of the rational differential form being pushed forward.
	\item If \smash{$\underline{\omega}^{I}_{J}$} is a constant with respect to $z_1,\ldots,z_n$, then \smash{$\mathcal{I}_\ast\underline{\omega}^{I}_{J}=d\,\underline{\omega}^{I}_{J}$} where $d= |\mathcal{V}(\mathcal{I})|$.
\end{itemize}

\section{Pushforwards via Companion Matrices}\label{sec:comp-matrices}

In this section we introduce the machinery of companion matrices \cite{sturmfels2002solving} and show how they can be used to compute \eqref{eq:pf-coefficients-pushforward} without needing to determine $\mathcal{V}(\mathcal{I})$. This method has already appeared in the scattering amplitudes literature \cite{Huang:2015yka} (see also \cite{Cardona:2015eba,Cardona:2015ouc}) where it was used to compute scattering amplitudes from moduli space integrals on the support of the scattering equations.

Recall that for generic $a$-variables, $\mathcal{I}=\langle f_1,\ldots,f_n\rangle\subseteq \mathbb{C}(\bm{a})[\bm{z}]$ is a zero-dimensional ideal and $Q=\mathbb{C}(\bm{a})[\bm{z}]/\mathcal{I}$ is a $d$-dimensional vector space over $\mathbb{C}(\bm{a})$. For each $i\in[n]$, multiplication by $z_i$ induces an endomorphism of $Q$
\begin{align}\label{eq:cm-endomorphism}
	\text{Times}_{z_i}:Q\to Q, \qquad f\mapsto z_if\,,
\end{align}
which can be represented by a matrix $T_i\in\mathsf{Mat}_d(\mathbb{C}(\bm{a}))$ in the standard basis $\mathcal{B}=\{e_\alpha\}_{\alpha=1}^{d}$. The components of $(T_i)_{\alpha\beta}$ are defined via
\begin{align}\label{eq:cm-matrix-components}
	(T_i)_{\alpha\beta}e_\beta:=\overbar{z_i e_\alpha}^{\mathcal{G}}  \,.
\end{align}
We call $T_i$ the \emph{i\textsuperscript{th} companion matrix} of $\mathcal{I}$. By definition, the companion matrices of $\mathcal{I}$ mutually commute and, hence, generate a commutative subalgebra of $\mathsf{Mat}_d(\mathbb{C}(\bm{a}))$ isomorphic to $Q$:
\begin{align}\label{eq:cm-isomorphism}
	\mathbb{C}(\bm{a})[T_1,\ldots,T_n]\cong Q,\qquad T_i\mapsto z_i\,.
\end{align}
The utility of companion matrices is expressed through \nameref{thm:stickelberger} which asserts that the complex affine variety $\mathcal{V}(\mathcal{I})$ is precisely the set of vectors $\bm{\lambda}=(\lambda_1,\ldots,\lambda_n)$ of simultaneous eigenvalues of the companion matrices of $\mathcal{I}$:
	\begin{align}
		\mathcal{V}(\mathcal{I}) &= \left\{\bm{\lambda}\in\overbar{\mathbb{C}(\bm{a})}^{n}\,|\,\exists_{\bm{v}\in\overbar{\mathbb{C}(\bm{a})}^{d}\setminus\{\bm{0}\}}\forall_{i\in[n]} : T_i\cdot \bm{v} = \lambda_i\,\bm{v}\right\}.
	\end{align}
It is a known fact that the companion matrices of $\mathcal{I}$ can be simultaneously diagonalized if and only if $\mathcal{I}$ is a radical ideal \cite{sturmfels2002solving}, and the latter is true precisely for generic $a$-variables as per the definition of $A_\text{gen}$.

To see how companion matrices can be used to evaluate \eqref{eq:pf-coefficients-pushforward}, note the following. Let $r\in\mathbb{C}(\bm{a})[\bm{z}]$ be any rational function.
Since the companion matrices of $\mathcal{I}$ mutually commute, the evaluation of $r$ on $\bm{T}=(T_1,\ldots,T_n)$ is well-defined. Moreover, since they are simultaneously diagonalizable, $r(\bm{T})$ is similar to a diagonal matrix whose diagonal entries are given by $\{r(\bm{\xi}):\bm{\xi}\in\mathcal{V}(\mathcal{I})\}$. Therefore
\begin{align}\label{eq:r-pushforward-cm}
	\mathcal{I}_\ast r\coloneqq \sum_{\bm{\xi}\in\mathcal{V}(\mathcal{I})}r(\bm{\xi})=\text{Tr}\left[r(\bm{T})\right].
\end{align}
Then, provided we can determine the companion matrices, the evaluation of \eqref{eq:pf-coefficients-pushforward} is reduced to a simple linear algebra calculation via \eqref{eq:r-pushforward-cm}. This result is demonstrated in the following example.
\begin{example}\label{ex:pf-cm}
	Recall that in \cref{ex:pf-formal} we showed that
	the coefficients defined in \eqref{eq:pf-coefficients} and \eqref{eq:pf-coefficients-pushforward} are given by
	\begin{align*}
		\underline{\omega}_{\{1,2\}}=0\,,\qquad\underline{\omega}_{\{1,3\}}=\frac{1}{2 a_1 a_3}\,,\qquad \underline{\omega}_{\{2,3\}}=\frac{z_2}{2 a_1 a_3 z_1}\,,
	\end{align*}
	and
	\begin{align*}
		\mathcal{I}_\ast\underline{\omega}_{\{1,2\}}=0\,,\qquad \mathcal{I}_\ast\underline{\omega}_{\{1,3\}} = \frac{1}{a_1a_3}\,,\qquad \mathcal{I}_\ast\underline{\omega}_{\{2,3\}} = -\frac{1}{a_2a_3}\,,
	\end{align*}
	respectively.
	Let us re-compute these pushforward coefficients using \eqref{eq:r-pushforward-cm}. Without loss of generality, assume grevlex-order where $z_1\succ z_2$. The Gr\"{o}bner basis for $\mathcal{I}$ is
	\begin{equation}
		\mathcal{G}=\left\{z_1+\frac{a_2}{a_1}z_2,z_2^2-\frac{1}{a_3}\right\},
		\end{equation}
	 and the standard basis for $Q$ is $\mathcal{B}=\{z_2,1\}$. The companion matrices of $\mathcal{I}$ are
	\begin{align}
		T_1=
		\begin{pmatrix}
			0 & -\frac{a_2}{a_1 a_3} \\
			-\frac{a_2}{a_1} & 0 \\
		\end{pmatrix},\qquad
		T_2=
		\begin{pmatrix}
			0 & \frac{1}{a_3} \\
			1 & 0 \\
		\end{pmatrix}
		.
	\end{align}
	Applying \eqref{eq:r-pushforward-cm} one obtains
	\begin{align}
		\mathcal{I}_\ast\underline{\omega}_{\{1,2\}}&=\text{Tr}[\underline{\omega}_{\{1,2\}}(T_1,T_2)] = \vphantom{\frac11}0\,,\\
		\mathcal{I}_\ast\underline{\omega}_{\{1,3\}}&=\text{Tr}[\underline{\omega}_{\{1,3\}}(T_1,T_2)] = \frac{1}{2a_1a_3}\text{Tr}[\mathbbm{1}_2] = \frac{1}{a_1a_3}\,,\\
		\mathcal{I}_\ast\underline{\omega}_{\{2,3\}}&=\text{Tr}[\underline{\omega}_{\{2,3\}}(T_1,T_2)] = \frac{1}{2 a_1 a_3}\text{Tr}[T_1^{-1}\cdot T_2] = -\frac{1}{a_2a_3}\,,
	\end{align}
	in agreement with the result above.
\end{example}

\section{Pushforwards via Derivatives of Companion Matrices}\label{sec:alternative-companion-matrices}

In the previous section we explained how \eqref{eq:pf-coefficients-pushforward}  can be computed by evaluating \eqref{eq:pf-coefficients} on companion matrices and taking the trace. One disadvantage of this method is that even for a relatively simple rational function $\underline{\omega}(\bm{z})$, the corresponding rational functions $\underline{\omega}^{I}_{J}$ given in \eqref{eq:pf-coefficients} can be highly complicated as a consequence of the multiplying $p\times p$ minor. Their complexity makes evaluating \eqref{eq:pf-coefficients} on companion matrices a non-trivial and time-consuming task.

In this section we present an alternative approach for computing the pushforward of rational differential forms via companion matrices which avoids needing to evaluate complicated minors on companion matrices. This method requires knowledge of the derivatives of companion matrices, and we present a novel algorithm for efficiently computing these partial derivatives for numerical data.

In \eqref{eq:pf-form-chain-rule}, we used the chain-rule to write the pushforward $\mathcal{I}_\ast\omega$ as
\begin{align*}
	\mathcal{I}_\ast\omega=\sum_{J\in \binom{[m]}{p}}\left[\sum_{\bm{\xi}\in\mathcal{V}(\mathcal{I})}\underline{\omega}(\bm{\xi})\bigg|\frac{\partial\bm{\xi}}{\partial\bm{a}}\bigg|^{I}_{J}\right]\bigwedge_{j\in J}\mathrm{d}a_{j}\,,
\end{align*}
Remarkably, the rational function prefactors multiplying \smash{$\bigwedge_{j\in J}\mathrm{d}a_{j}$} can be computed by uplifting this expression to one involving companion matrices in the following natural manner:
\begin{align}\label{eq:alt-coefficients-pushforward}
	\sum_{\bm{\xi}\in\mathcal{V}(\mathcal{I})}\underline{\omega}(\bm{\xi})\bigg|\frac{\partial\bm{\xi}}{\partial\bm{a}}\bigg|^{I}_{J} = \mathrm{Tr}\left[\underline{\omega}(\bm{T})\sum_{\mathclap{\sigma\in S_p}}\mathrm{sgn}(\sigma)\frac{\partial T_{i_{\sigma(1)}}}{\partial a_{j_1}}\cdots\frac{\partial T_{i_{\sigma(p)}}}{\partial a_{j_p}}\right] \eqqcolon \mathcal{I}_\ast\underline{\omega}^{I}_{J}.
\end{align}
This formula is non-trivial since the partial derivatives of companion matrices are not guaranteed to simultaneously commute, and in general they do not. Before giving a proof of this formula, let us see it applied to our running example.

\begin{example}\label{ex:pf-alt}
	Recall that in \cref{ex:pf-formal} we considered the pushforward of
	\begin{align*}
		\omega=\mathrm{d}\log z_1\wedge \mathrm{d}\log z_2 = \underline{\omega}(\bm{z})\,\mathrm{d}z_1\wedge \mathrm{d}z_2\,,\qquad\text{where}\qquad\underline{\omega}(\bm{z})=\frac{1}{z_1z_2}\,.
	\end{align*}
	In \cref{ex:pf-cm} we confirmed that the rational function prefactors of the pushforward as defined in \eqref{eq:pf-coefficients-pushforward} were given by
	\begin{align*}
		\mathcal{I}_\ast\underline{\omega}_{\{1,2\}}=0\,,\qquad \mathcal{I}_\ast\underline{\omega}_{\{1,3\}} = \frac{1}{a_1a_3}\,,\qquad \mathcal{I}_\ast\underline{\omega}_{\{2,3\}} = -\frac{1}{a_2a_3}\,,
	\end{align*}
	where we used the companion matrices
	\begin{align*}
		T_1=
		\begin{pmatrix}
			0 & -\frac{a_2}{a_1 a_3} \\
			-\frac{a_2}{a_1} & 0 \\
		\end{pmatrix},\qquad
		T_2=
		\begin{pmatrix}
			0 & \frac{1}{a_3} \\
			1 & 0 \\
		\end{pmatrix}
		.
	\end{align*}
	The derivatives of these matrices are given by
	\begin{alignat}{3}
		\frac{\partial T_1}{\partial a_1} &= \begin{pmatrix}
			0 & \frac{a_2}{a_1^2a_3} \\
			\frac{a_2}{a_1^2} & 0
		\end{pmatrix}, \qquad& \frac{\partial T_1}{\partial a_2} &= \begin{pmatrix}
			0 & -\frac{1}{a_1a_3} \\
			-\frac{1}{a_1} & 0
		\end{pmatrix}, \qquad& \frac{\partial T_1}{\partial a_3} &=
		\begin{pmatrix}
			0 & \frac{a_2}{a_1a_3^2} \\
			0 & 0
		\end{pmatrix},\\
		\frac{\partial T_2}{\partial a_1} &=
		0_{2\times 2}, \qquad& \frac{\partial T_2}{\partial a_2} &=
			0_{2\times 2}, \qquad& \frac{\partial T_2}{\partial a_3} &=
		\begin{pmatrix}
			0 & -\frac{1}{a_3^2} \\
			0 & 0
		\end{pmatrix}.
	\end{alignat}
	Using \eqref{eq:alt-coefficients-pushforward}, one finds that
	\begin{align}
		\mathcal{I}_\ast \omega_{\{1,2\}} &= \mathrm{Tr}\left[\underline{\omega}(\bm{T})\left(\frac{\partial T_1}{\partial a_1}\frac{\partial T_2}{\partial a_2}-\frac{\partial T_2}{\partial a_1}\frac{\partial T_1}{\partial a_2}\right)\right] = 0\,,\\
		\mathcal{I}_\ast \omega_{\{1,3\}} &= \mathrm{Tr}\left[\underline{\omega}(\bm{T})\left(\frac{\partial T_1}{\partial a_1}\frac{\partial T_2}{\partial a_3}-\frac{\partial T_2}{\partial a_1}\frac{\partial T_1}{\partial a_3}\right)\right] = \mathrm{Tr}\left[(T_1 T_2)^{-1}\frac{\partial T_1}{\partial a_1}\frac{\partial T_2}{\partial a_3}\right] = \frac{1}{a_1a_3}\,,\\
		\mathcal{I}_\ast \omega_{\{2,3\}} &= \mathrm{Tr}\left[\underline{\omega}(\bm{T})\left(\frac{\partial T_1}{\partial a_2}\frac{\partial T_2}{\partial a_3}-\frac{\partial T_2}{\partial a_2}\frac{\partial T_1}{\partial a_3}\right)\right] = \mathrm{Tr}\left[(T_1 T_2)^{-1}\frac{\partial T_1}{\partial a_2}\frac{\partial T_2}{\partial a_3}\right] = - \frac{1}{a_2a_3}\,,
	\end{align}
	as expected.
\end{example}

\begin{proof}[Proof of \eqref{eq:alt-coefficients-pushforward}] Recall that
\begin{itemize}
	\item The companion matrices of $\mathcal{I}$ are assumed to be simultaneously diagonalizable, i.e.\ there exists an invertible matrix $P\in\mathsf{GL}_{d}(\overbar{\mathbb{C}(\bm{a})})$ such that each companion matrix $T_i$ can be written as \smash{$T_i=P D_i P^{-1}$} where $D_i$ is a diagonal matrix.
	\item The set of vectors of their simultaneous eigenvalues \smash{$\{\bm{\lambda}^{(\alpha)}\}_{\alpha=1}^{d}$} is precisely $\mathcal{V}(\mathcal{I})$ (as a consequence of \nameref{thm:stickelberger}) and, thus, the matrix components of $D_i$ are $(D_i)_{\alpha\beta}=\lambda^{(\alpha)}_i\delta_{\alpha\beta}$.
\end{itemize}
Consequently, it is sufficient to prove that
\begin{align}\label{eq:alt-proof-want}
	\mathrm{Tr}\left[\underline{\omega}(\bm{T})\sum_{\mathclap{\sigma\in S_p}}\mathrm{sgn}(\sigma)\frac{\partial T_{i_{\sigma(1)}}}{\partial a_{j_1}}\cdots\frac{\partial T_{i_{\sigma(p)}}}{\partial a_{j_p}}\right]= \mathrm{Tr}\left[\underline{\omega}(\bm{D})\sum_{\mathclap{\sigma\in S_p}}\mathrm{sgn}(\sigma)\frac{\partial D_{i_{\sigma(1)}}}{\partial a_{j_1}}\cdots\frac{\partial D_{i_{\sigma(p)}}}{\partial a_{j_p}}\right].
\end{align}
The pairwise commutativity of companion matrices implies that
\begin{align}\label{eq:alt-proof-rational-function}
	\underline{\omega}(\bm{T})=P \underline{\omega}(\bm{D}) P^{-1}\,.
\end{align}
Differentiating \smash{$T_i=P D_i P^{-1}$} with respect to $a_j$ yields
\begin{align}\label{eq:alt-proof-derivatives}
	\frac{\partial T_i}{\partial a_j} = P \left(\frac{\partial D_i}{\partial a_j}+ [\Gamma_j,D_i]\right) P^{-1}\,,\qquad\text{where}\qquad\Gamma_j\coloneqq P^{-1} \frac{\partial P}{\partial a_j}\,.
\end{align}
Substituting \eqref{eq:alt-proof-rational-function} and \eqref{eq:alt-proof-derivatives} into the left-hand side (LHS) of \eqref{eq:alt-proof-want} produces
\begin{align}\label{eq:alt-proof-want-LHS}
	\mathrm{LHS}\eqref{eq:alt-proof-want} = \sum_{r=0}^{p}\sum_{K\in\binom{[p]}{r}}\sum_{\sigma\in S_p}\mathrm{sgn}(\sigma)\,\mathrm{Tr}^{I}_{J}(K;\sigma)\,,
\end{align}
where
\begin{align}\label{eq:alt-proof-TR}
	\mathrm{Tr}^{I}_{J}(K;\sigma)\coloneqq\mathrm{Tr}\left[\underline{\omega}(\bm{D}) \prod_{k=1}^{p}
	\left\{\begin{array}{ll}
		\frac{\partial}{\partial a_{j_k}}D_{i_{\sigma(k)}}&\text{if}~k\notin K\\
		{[\Gamma_{j_k},D_{i_{\sigma(k)}}]}&\text{if}~k\in K
	\end{array}\right\}
	\right].
\end{align}
The contribution to \eqref{eq:alt-proof-want-LHS} from $r=0$, for which $K=\emptyset$, is clearly the right-hand side of \eqref{eq:alt-proof-want}. Therefore, we need to show that the contributions for $r>0$ vanish. To this end, let $r>0$ and fix \smash{$K=\{k_1,\ldots,k_{r}\}\in\binom{[p]}{r}$}. In matrix components, \eqref{eq:alt-proof-TR} is given by
\begin{align}\label{eq:alt-proof-TR-components}
	\begin{split}
		&\mathrm{Tr}^{I}_{J}(K;\sigma)= \sum_{{1\le\alpha_1,\ldots,\alpha_{r}\le d}}\, \underline{\omega}\big(\bm{\lambda}^{(\alpha_1)}\big)\\
		&\times \left[\prod_{s=1}^{r}
		\left[\prod_{k=k_{s-1}+1}^{k_{s}-1}
		\frac{\partial}{\partial a_{j_k}}\lambda_{i_{\sigma(k)}}^{(\alpha_s)}\right]
		\big(\Gamma_{j_{k_s}}\big)_{\alpha_s\alpha_{s+1}}
		\left(\lambda_{i_{\sigma(k_s)}}^{(\alpha_{s+1})}-\lambda_{i_{\sigma(k_s)}}^{(\alpha_{s})}\right)\right]\left[\prod_{k=k_{r}+1}^{p}
		\frac{\partial}{\partial a_{j_k}}\lambda_{i_{\sigma(k)}}^{(\alpha_1)}\right],
	\end{split}
\end{align}
where $k_0=0$ and $\alpha_{r+1}=\alpha_{1}$, and we have written the sum over the matrix components $\alpha_1,\ldots,\alpha_{r}$ explicitly. We now distinguish two cases.
\begin{itemize}
	\item If $r<p$, then let $K'=[p]\setminus K = \{k'_1,\ldots,k'_{p-r}\}$. In this case we can always re-express the sum over permutations in $S_p$ as
	\begin{align}\label{eq:alt-proof-permutations}
\sum_{\sigma\in S_p} = \sum_{L'\in\binom{[p]}{p-r}}\sum_{\pi\in S_{p-r}}\sum_{\sigma\in S_{p}(K',L';\pi)}\,,
	\end{align}
	where the first sum is over $L'=\{l'_1,\ldots,l'_{p-r}\}\in\binom{[p]}{p-k}$ and the third sum is over
	\begin{align}
		S_{p}(K',L';\pi)\coloneqq\left\{\sigma\in S_p\,| \,\forall_{\gamma\in[p-r]}:\, \sigma(k'_\gamma)=l'_{\pi(\gamma)}\right\},
	\end{align}
	i.e.\ the set of $r!$ permutations in $S_p$ which map $K'$ to $L'$ in a fixed manner according to $\pi$. For each $L'=\{l'_1,\ldots,l'_{p-r}\}\in\binom{[p]}{p-k}$ and $\pi\in S_{p-r}$, it is easy to show that
	\begin{align}
			\sum_{\sigma\in S_p(K',L';\pi)}\mathrm{sgn}(\sigma)\prod_{s=1}^{r}\left(\lambda_{i_{\sigma(k_s)}}^{(\alpha_{s+1})}-\lambda_{i_{\sigma(k_s)}}^{(\alpha_{s})}\right) = 0\,,
	\end{align} 
	for any choice of $1\le\alpha_1,\ldots,\alpha_{r}\le d$. Thus, combining \eqref{eq:alt-proof-TR-components} with \eqref{eq:alt-proof-permutations}, we obtain \smash{$\sum_{\sigma\in S_p}\mathrm{Tr}^{I}_{J}(K;\sigma)=0$}.
	\item If $r=p$, then there are no longer derivative terms in \eqref{eq:alt-proof-TR-components}. Then since
	\begin{align}
		\sum_{\sigma\in S_p}\mathrm{sgn}(\sigma)\prod_{s=1}^{p}\left(\lambda_{i_{\sigma(k_s)}}^{(\alpha_{s+1})}-\lambda_{i_{\sigma(k_s)}}^{(\alpha_{s})}\right) = 0\,,
	\end{align}
	for fixed $1\le\alpha_1,\ldots,\alpha_{r}\le d$, it again follows that \smash{$\sum_{\sigma\in S_p}\mathrm{Tr}^{I}_{J}(K;\sigma)=0$}.
\end{itemize}
Altogether we have that \eqref{eq:alt-proof-want-LHS} vanishes except when $r=0$ and $K=\emptyset$. This concludes our proof of \eqref{eq:alt-coefficients-pushforward}.
\end{proof}

In many cases, evaluating \eqref{eq:alt-coefficients-pushforward} is more efficient than evaluating \eqref{eq:pf-coefficients-pushforward} using companion matrices, provided one can determine the companion matrices for generic $a$-variables. However, in some cases it is not tractable to compute companion matrices in full generality, but only for numerical values of $a$-variables. How then can one determine the derivatives of companion matrices with respect to $a$-variables for numerical values? It turns out that, using lex ordering, we can solve this problem using the following algorithm.

To explain our algorithm, fix $1\le j\le m$ and let $\bm{a}\in A_\text{gen}$ be a generic $a$-variable.
Let
\begin{align}
	S\coloneqq\mathbb{C}[z_1,\ldots,z_n]\,,
\end{align}
and
\begin{align}
	S_j\coloneqq\mathbb{C}\left[\frac{\partial z_1}{\partial a_j},\ldots,\frac{\partial z_n}{\partial a_j},z_1,\ldots,z_n\right],
\end{align}
be polynomial rings, where \smash{${\partial z_1}/{\partial a_j},\ldots,{\partial z_n}/{\partial a_j}$} are regarded as formal variables. Recall that $\mathcal{I}(\bm{a})$ is an ideal in $S$. Define the ideals
\begin{align}
	\frac{d \mathcal{I}}{d a_j}(\bm{a}) \coloneqq \left\langle \frac{df_1}{da_j}(\bullet;\bm{a}),\ldots,\frac{df_n}{da_j}(\bullet;\bm{a})\right\rangle = \left\langle \frac{\partial f_1}{\partial a_j} + \frac{\partial f_1}{\partial z_i}\frac{\partial z_i}{\partial a_j},\ldots,\frac{\partial f_n}{\partial a_j} + \frac{\partial f_n}{\partial z_i}\frac{\partial z_i}{\partial a_j}\right\rangle \subseteq S_j\,,
\end{align}
where again \smash{${\partial z_1}/{\partial a_j},\ldots,{\partial z_n}/{\partial a_j}$} are regarded as formal variables, and
\begin{align}
	\mathcal{I}_j(\bm{a})\coloneqq \mathcal{I}(\bm{a})+ \frac{d \mathcal{I}}{d a_j}(\bm{a})\subseteq S_j\,,
\end{align}
which is generated by the generators of $\mathcal{I}(\bm{a})$ and \smash{${d\mathcal{I}}/{da_j}\,(\bm{a})$}. Let $\mathcal{G}(\bm{a})$ and $\mathcal{G}_j(\bm{a})$ be the Gr\"{o}bner bases for $\mathcal{I}(\bm{a})$ and $\mathcal{I}_j(\bm{a})$, respectively, using lex ordering with ${\partial z_1}/{\partial a_j}\succ\cdots\succ{\partial z_n}/{\partial a_j}\succ z_1\succ\cdots\succ z_n$. Furthermore, let $\mathcal{B}(\bm{a})$ and $\mathcal{B}_j(\bm{a})$ be the standard bases for $S/\mathcal{I}(\bm{a})$ and $S_j/\mathcal{I}_j(\bm{a})$, respectively. We will assume that $\mathcal{B}_j(\bm{a})=\mathcal{B}(\bm{a})$. We do not know what the necessary nor the sufficient conditions for this assumption are, but we note that it is true for all the examples which we consider in this paper.

In this setting, and with the above assumption, the $i$\textsuperscript{th} companion matrix $T_i(\bm{a})$ (evaluated at $\bm{a}$) of $\mathcal{I}(\bm{a})$ can be computed via
\begin{align}\label{eq:alt-cm-components}
	\overbar{z_i e_\alpha}^{\mathcal{G}_j(\bm{a})} = T_i(\bm{a})_{\alpha\beta}e_{\beta}\,,
\end{align}
and its partial derivative \smash{$\frac{\partial T_i}{\partial a_j}(\bm{a})$} (evaluated at $\bm{a}$) with respect to $a_j$ can be calculated using
\begin{align}\label{eq:alt-cm-pd-components}
	\overbar{\frac{\partial(z_i e_\alpha)}{\partial a_j} - T_i(\bm{a})_{\alpha\beta}\frac{\partial e_\beta}{\partial a_j}}^{\mathcal{G}_j(\bm{a})} = \frac{\partial T_i(\bm{a})_{\alpha\beta}}{\partial a_j}e_\beta\,.
\end{align}
Notice that in both \eqref{eq:alt-cm-components} and \eqref{eq:alt-cm-pd-components}, the objects on the left-hand side that are being divided by $\mathcal{G}_{j}(\bm{a})$ are polynomials in \smash{$S_j$} while the polynomials on the right-hand side are polynomials in $S$.

To understand why \eqref{eq:alt-cm-components} is true, notice that each generator \smash{${df_\ell(\bullet;\bm{a})}/{da_j}$} of \smash{${d \mathcal{I}(\bm{a})}/{d a_j}$} defines a non-constant linear function in the formal variables \smash{${\partial z_1}/{\partial a_j},\ldots,{\partial z_n}/{\partial a_j}$}. Consequently, the \emph{$n$\textsuperscript{th} elimination ideal} of $\mathcal{I}_j(\bm{a})$ is
\begin{align}
\mathcal{I}_j(\bm{a})\cap S = \mathcal{I}(\bm{a})\,,
\end{align}
and by the \nameref{thm:elimination} we have that $\mathcal{G}_j(\bm{a})\cap S = \mathcal{G}(\bm{a})$. (We restricted our attention to lex ordering in order to use this result.) Furthermore, given our assumption that $\mathcal{B}_j(\bm{a})=\mathcal{B}(\bm{a})$, it follows that $z_i e_\alpha$ is a monomial in $S$ and, hence, division by $\mathcal{G}_j(\bm{a})$ is identical to division by $\mathcal{G}(\bm{a})$. Consequently, \eqref{eq:alt-cm-components} does indeed compute the $i$\textsuperscript{th} companion matrix of $\mathcal{I}(\bm{a})$.

What about \eqref{eq:alt-cm-pd-components}? From \eqref{eq:alt-cm-components} we have that
\begin{align}
	z_i e_\alpha \simeq T_i(\bm{a})_{\alpha\beta}e_{\beta}\,,
\end{align}
where $\simeq$ means equality modulo division by $\mathcal{G}_j(\bm{a})$. Differentiating both sides with respect to $a_j$ (and regarding $z$-variables as implicit functions of $a_j$) produces
\begin{align}
	\frac{\partial(z_i e_\alpha)}{\partial a_j} \simeq \frac{T_i(\bm{a})_{\alpha\beta}}{\partial a_j}e_{\beta} + T_i(\bm{a})_{\alpha\beta}\frac{\partial e_\beta}{\partial a_j}\,,
\end{align}
from which \eqref{eq:alt-cm-pd-components} follows.

To demonstrate the mechanics of this algorithm, let us return to our running example.
\begin{example} Recall from \cref{ex:pf-alt} that the derivatives of the companion matrices\footnote{For this example, the companion matrices for grevlex-order and lexicographic order are the same.} in \cref{ex:pf-cm} are given by
		\begin{alignat*}{3}
		\frac{\partial T_1}{\partial a_1} &= \begin{pmatrix}
			0 & \frac{a_2}{a_1^2a_3} \\
			\frac{a_2}{a_1^2} & 0
		\end{pmatrix}, \qquad& \frac{\partial T_1}{\partial a_2} &= \begin{pmatrix}
			0 & -\frac{1}{a_1a_3} \\
			-\frac{1}{a_1} & 0
		\end{pmatrix}, \qquad& \frac{\partial T_1}{\partial a_3} &=
		\begin{pmatrix}
			0 & \frac{a_2}{a_1a_3^2} \\
			0 & 0
		\end{pmatrix},\\
		\frac{\partial T_2}{\partial a_1} &= 0_{2\times 2}, \qquad& \frac{\partial T_2}{\partial a_2} &=
		0_{2\times 2}, \qquad& \frac{\partial T_2}{\partial a_3} &=
		\begin{pmatrix}
			0 & -\frac{1}{a_3^2} \\
			0 & 0
		\end{pmatrix}.
	\end{alignat*}
	Let us recompute these using \eqref{eq:alt-cm-pd-components}. Starting with derivatives with respect to $a_1$, the relevant ideal is
	\begin{align}
		\mathcal{I}_1 = \left\langle f_1,f_2,\frac{df_1}{da_1},\frac{df_2}{da_1}\right\rangle =  \left\langle a_1z_1+a_2z_2,a_3z_2^2-1,z_1+a_1\frac{\partial z_1}{\partial a_1}+a_2\frac{\partial z_2}{\partial a_1},2 a_3 z_2\frac{\partial z_2}{\partial a_1}\right\rangle,
	\end{align}
	and its Gr\"{o}bner basis with respect to lex ordering is
	\begin{align}
		\mathcal{G}_1=\left\{\frac{\partial z_1}{\partial a_1} - \frac{a_2}{a_1^2}z_2,\frac{\partial z_2}{\partial a_1},z_1+\frac{a_2}{a_1}z_2,z_2^2-\frac{1}{a_3}
		\right\}.
	\end{align}
	Clearly the standard basis for $\mathcal{I}_1$ is $\mathcal{B}_1=\{z_2,1\}$ which is precisely the standard basis $\mathcal{B}$ for $\mathcal{I}$. Let \smash{$\bm{e}=(z_2,1)^T$}. It is easy to show that
	\begin{align}
	\frac{\partial(z_1 \bm{e})}{\partial a_j} - T_1\cdot\frac{\partial \bm{e}}{\partial a_j} &= \left(z_2\frac{\partial z_1}{\partial a_1}+z_1\frac{\partial z_2}{\partial a_1},\frac{\partial z_1}{\partial a_1} + \frac{a_2}{a_1}\frac{\partial z_2}{\partial a_1}\right)^T \simeq \left(\frac{a_2}{a_1^2 a_3},\frac{a_2 z_2}{a_1^2}\right)^T, \\
	\frac{\partial(z_2 \bm{e})}{\partial a_j} - T_2\cdot\frac{\partial \bm{e}}{\partial a_j} &= \left(2z_2\frac{\partial z_2}{\partial a_1},0\right)^T \simeq (0,0)^T\,,
	\end{align}
	from which we read off the formulae for the derivatives of the companion matrices with respect to $a_1$
	\begin{align}
\frac{\partial T_1}{\partial a_1}= \begin{pmatrix}
	0 & \frac{a_2}{a_1^2a_3} \\
	\frac{a_2}{a_1^2} & 0
\end{pmatrix}, \qquad\frac{\partial T_2}{\partial a_1}= 0_{2\times 2}\,,
	\end{align}
	in agreement with what we found in \cref{ex:pf-alt}. A similar calculation confirms the other derivatives.
\end{example}

\section{Pushforwards via the Global Duality of Residues}\label{sec:global-residue}

In this section we consider a different method for calculating \eqref{eq:pf-coefficients-pushforward} which uses global residues and a special basis for the quotient ring $Q=\mathbb{C}(\bm{a})[\bm{z}]/\mathcal{I}$ which is dual to the standard basis $\mathcal{B}$. This method has also already appeared in the context of the scattering equations \cite{Sogaard:2015dba}. It has some computational advantages over the previous method as it does not necessitate the computation of companion matrices which grow very quickly in size. For instance, for the $n$-particle CHY scattering equations, the companion matrices are of size $(n-3)!\times(n-3)!$.

\paragraph{Global Residues.}

Suppose that we wish to sum a rational function $r\in\mathbb{C}(\bm{a})\allowbreak(\bm{z})$ over all points $\bm{\xi}\in \mathcal{V}(\mathcal{I})$. We will later specialise to the case where \smash{$r=\underline{\omega}^{\{1,\ldots,n\}}_{\{j_1,\ldots,j_n\}}$} where some simplifications occur. Following \cite{Sogaard:2015dba}, let us uplift this rational function to a rational differential form with simple poles on $\mathcal{V}(\mathcal{I})$ as follows:
\begin{align}\label{eq:gr-form}
	\Omega(r)\coloneqq r(\bm{z}) \frac{\mathrm{d} z_1\wedge\cdots\wedge\mathrm{d}z_n }{f_1(\bm{z})\cdots f_n(\bm{z})}\,.
\end{align}
For each $\bm{\xi}\in\mathcal{V}(\mathcal{I})$, we define the \emph{local residue of $\Omega(r)$ at $\bm{\xi}$} through the multi-dimensional contour integral
\begin{align}\label{eq:gr-local-residue}
	\mathrm{Res}_{\bm{\xi}}(\Omega(r))\coloneqq \frac{1}{(2\pi i)^n}\oint_{\Gamma_{\bm{\xi}}} \Omega(r)\,,
\end{align}
where $\Gamma_{\bm{\xi}}$ is some suitably defined contour around the pole $\bm{\xi}$, a definition of which can be found in \cite{Sogaard:2015dba}. We further define the \emph{global residue of $r$} as
\begin{align}\label{eq:gr-global-residue}
	\mathrm{Res}(r)\coloneqq \sum_{\bm{\xi}\in\mathcal{V}(\mathcal{I})}\mathrm{Res}_{\bm{\xi}}(\Omega(r))\,.
\end{align}
Using the multi-dimensional form of Cauchy's theorem, we have that
\begin{align}\label{eq:gr-cauchy}
	\mathrm{Res}_{\bm{\xi}}(\Omega(r))=r(\bm{\xi})\bigg|\frac{\partial\bm{f}}{\partial\bm{z}}(\bm{\xi})\bigg|^{-1},
\end{align}
where \smash{$\big|{\partial\bm{f}}/{\partial\bm{z}}\big|$} is the determinant of the Jacobian matrix \smash{${\partial\bm{f}}/{\partial\bm{z}}$}. Hence, our quantity of interest can be expressed as the global residue of \smash{$\big|{\partial\bm{f}}/{\partial\bm{z}}\big|\,r$}:
\begin{align}\label{eq:r-pushforward-gr}
	\mathcal{I}_\ast r\coloneqq\sum_{\bm{\xi}\in\mathcal{V}(\mathcal{I})} r(\bm{\xi})=\mathrm{Res}\bigg(\bigg|\frac{\partial\bm{f}}{\partial\bm{z}}\bigg| r\bigg)\,.
\end{align}

\paragraph{Duality of Global Residues and the Dual Basis.} Following \cite{Sogaard:2015dba}, we can evaluate \eqref{eq:r-pushforward-gr} using the so-called global duality theorem which requires us to momentarily restrict our attention to polynomials. We will later extend our analysis to include rational functions.

The following fact is a defining characteristic of Gr\"{o}bner bases: the remainder of a polynomial $p\in\mathbb{C}(\bm{a})[\bm{z}]$ on division by $\mathcal{G}$, denoted by $\overbar{p}^{\mathcal{G}}$, is unique. Moreover, when a polynomial is evaluated on any point $\bm{\xi}\in \mathcal{V}(\mathcal{I})$, only its remainder survives:
\begin{align}
	p(\bm{\xi}) = \overbar{p}^{\mathcal{G}}(\bm{\xi})= \sum_{\alpha=1}^{d}{p}_{\alpha}e_\alpha(\bm{\xi}),\qquad\forall\, \bm{\xi}\in\mathcal{V}(\mathcal{I}) \,,
\end{align}
where \smash{$\mathcal{B}=\{e_\alpha\}_{\alpha=1}^{d}$} is the standard basis for $Q$ and ${p}_{\alpha}\in\mathbb{C}(\bm{a})$ are the components of $p$ modulo $\mathcal{G}$ in this basis.
Consequently, it is natural to address the problem of computing \smash{$\mathcal{I}_\ast p = \sum_{\bm{\xi}\in\mathcal{V}(\mathcal{I})} p(\bm{\xi})$} directly in $Q$. To this end, define the following symmetric inner product on $Q$:
\begin{align}\label{eq:gr-inner-product}
	\langle \bullet,\bullet\rangle \colon Q\times Q \to \mathbb{C}(\bm{a}), \quad (p_1,p_2)\mapsto \langle p_1,p_2\rangle \coloneqq \mathrm{Res}(p_1\, p_2)\,.
\end{align}
The \nameref{thm:global-duality} tells us that $\langle\bullet,\bullet\rangle$ is a non-degenerate inner-product on $Q$ which implies the existence of a \emph{dual basis} \smash{$\mathcal{B}^\vee=\mathcal{B}^\vee_\prec(\mathcal{I})\coloneqq\{\Delta_\alpha\}_{\alpha=1}^{d}$} for $Q$ such that
\begin{align}
	\langle e_\alpha,\Delta_\beta\rangle=\delta_{\alpha\beta}\,.
\end{align}
Since $1\in Q$, it is always possible to express $1$ in the dual basis as
\begin{align}
1\eqqcolon\sum_{\alpha=1}^{d} \mu_\alpha\Delta_\alpha\,,
\end{align}
where $\mu_\alpha\in\mathbb{C}(a_1,\ldots,a_m)$ are the components of $1$ in this basis.

Finally, the global residue of \smash{$\big|{\partial\bm{f}}/{\partial\bm{z}}\big|\,p$} can be calculated by decomposing the remainder of \smash{$\big|{\partial\bm{f}}/{\partial\bm{z}}\big|\,p$} modulo $\mathcal{G}$ with respect to $\mathcal{B}$
\begin{align}
	\overbar{\bigg(\bigg|\frac{\partial\bm{f}}{\partial\bm{z}}\bigg|p\bigg)}^{\mathcal{G}}\eqqcolon\sum_{\alpha=1}^{d} {\bigg(\bigg|\frac{\partial\bm{f}}{\partial\bm{z}}\bigg|p\bigg)}_\alpha e_\alpha\,,
\end{align}
and taking its inner product with $1$
\begin{align}\label{eq:p-global-residue}
	\mathrm{Res}\bigg(\bigg|\frac{\partial\bm{f}}{\partial\bm{z}}\bigg|p\bigg) = \left\langle \overbar{\bigg(\bigg|\frac{\partial\bm{f}}{\partial\bm{z}}\bigg|p\bigg)}^{\mathcal{G}},1 \right\rangle =\left\langle \sum_{\alpha=1}^{d} {\bigg(\bigg|\frac{\partial\bm{f}}{\partial\bm{z}}\bigg|p\bigg)}_\alpha e_\alpha, \sum_{\beta=1}^{d}\mu_\beta \Delta_\beta  \right\rangle
	= \sum_{\alpha=1}^{d}{\bigg(\bigg|\frac{\partial\bm{f}}{\partial\bm{z}}\bigg|p\bigg)}_\alpha\mu_\alpha\,.
\end{align}
Therefore, once the universal coefficients $\mu_\alpha$ are determined, $\mathcal{I}_\ast p$ can be computed as
\begin{align}\label{eq:p-pushforward-gr-dual}
	\mathcal{I}_\ast p = \mathrm{Res}\bigg(\bigg|\frac{\partial\bm{f}}{\partial\bm{z}}\bigg|p\bigg)=
	\sum_{\alpha=1}^{d}{\bigg(\bigg|\frac{\partial\bm{f}}{\partial\bm{z}}\bigg|p\bigg)}_\alpha\mu_\alpha\,.
\end{align}

\emph{What about rational functions?} Naively, it seems that this method for calculating the global residue of polynomials cannot extend to rational functions due to the presence of denominators. However, for most rational functions we can find a polynomial which will yield the same result when summed over $\mathcal{V}(\mathcal{I})$. This is achieved by replacing each denominator by its \textit{polynomial inverse} in $Q$. Explicitly, let us take a rational function $r(\bm{z})={p(\bm{z})}/{q(\bm{z})}$ where $p$ and $q$ are polynomials, and suppose $q$ does not vanish for any point in $\mathcal{V}(\mathcal{I})$. Since $f_1,\ldots,f_n,q$ have no common zeroes, $1\in\langle f_1,\ldots,f_n,q\rangle$ by \nameref{thm:nullstellensatz} and so we can then find polynomials $\tilde{f}_1,\ldots,\tilde{f}_n,q_\text{inv}$ such that
\begin{align}
	1 = \tilde{f}_1f_1+\ldots+\tilde{f}_nf_n + q_\text{inv}\,q\implies \overbar{q_\text{inv}\,q}^{\mathcal{G}} = 1\,.
\end{align}
Once  the polynomial inverse $q_\text{inv}$ is found, one can then compute $\mathcal{I}_\ast r$ as
\begin{align}\label{eq:gr-final-r}
	\mathcal{I}_\ast r = \mathrm{Res}\bigg(\bigg|\frac{\partial\bm{f}}{\partial\bm{z}}\bigg|p\,q_\text{inv}\bigg)= \sum_{\alpha=1}^{d}{\bigg(\bigg|\frac{\partial\bm{f}}{\partial\bm{z}}\bigg|p\,q_\text{inv}\bigg)}_{\alpha} \mu_\alpha\,.
\end{align}

To summarise, we have reduced the problem of summing a rational function over the points in $\mathcal{V}(\mathcal{I})$ to finding a Gr\"obner basis $\mathcal{G}$, the standard basis $\mathcal{B}$, the dual basis $\mathcal{B}^\vee$ and polynomial inverses for denominators. For the sake of completeness, we now briefly outline how to calculate the dual basis and polynomial inverses, as detailed in \cite{Sogaard:2015dba}.

\paragraph{Calculating Dual Bases.} First compute the
Gr\"{o}bner basis $\mathcal{G}$ and the standard basis \smash{$\mathcal{B}=\{e_\alpha\}_{\alpha=1}^{d}$} where $\prec$ is either grlex or grevlex order. Then introduce some auxiliary variables $y_1,\ldots,y_n$ and define the \textit{Bezoutian matrix} $B$ for $\mathcal{I}$ with components
\begin{align}\label{eq:gr-bezout}
	B_{i,j}\coloneqq \frac{f_i(y_1,\ldots,y_{j-1},z_j,\ldots,z_n)-f_i(y_1,\ldots,y_{j},z_{j+1},\ldots,z_n)}{z_j-y_j}\,.
\end{align}
Let \smash{$\tilde{\mathcal{G}}\coloneqq \mathcal{G}\big|_{\bm{z}\to\bm{y}}$}. The remainder of $\det B$ on division by $\mathcal{G}\cup \tilde{\mathcal{G}}$ can be decomposed in the standard monomial basis as
\begin{align}
	\overbar{\det B}^{\mathcal{G}\cup \tilde{\mathcal{G}}} \eqqcolon \sum_{\alpha=1}^{d} {(\det B)}_{\alpha} e_\alpha\,,
\end{align}
where \smash{${(\det B)}_{\alpha}\in\mathbb{C}(\bm{a})[\bm{y}]$} and $\bm{y}=(y_1,\ldots,y_n)$.
We then find the dual basis $\mathcal{B}^\vee=\{\Delta_\alpha\}_{\alpha=1}^{d}$ with respect to the inner product $\langle \bullet,\bullet\rangle$ by the evaluating \smash{${(\det B)}_{\alpha}$} on $\bm{y}=\bm{z}$ \cite{cattani2005introduction}
\begin{align}\label{eq:bez-dual-basis}
\Delta_\alpha ={(\det B)}_{\alpha}(\bm{z})\,.
\end{align}

\begin{example}\label{ex:dual-basis}
	Recall from \cref{ex:pf-cm} that $\mathcal{G}= \{z_1+\frac{a_2}{a_1}z_2,z_2^2-\frac{1}{a_3}\}$ and $\mathcal{B}=\{e_1,e_2\}\coloneqq\{z_2,1\}$. Introducing auxiliary variables $y_1,y_2$ the Bezoutian matrix $B$ for $\mathcal{I}$ is calculated to be
	\begin{align}
		B=\begin{pmatrix}
			a_1 & a_2 \\
			0 & a_3(z_2+y_2)
		\end{pmatrix}.
	\end{align}
	The remainder of $\det B$ on division by $\mathcal{G}\cup\tilde{\mathcal{G}}$ expressed in terms of $\mathcal{B}$ is given by
	\begin{align}
		\overbar{\det B}^{\mathcal{G}\cup\tilde{\mathcal{G}}}=a_1a_3(z_2+y_2) = (a_1a_3)e_1+(a_1a_3y_2)e_2\,,
	\end{align}
	from which we read off the dual basis $\mathcal{B}^\vee=\{\Delta_1,\Delta_2\}\coloneqq \{a_1a_3,a_1a_3z_2\}$. Furthermore, since $\Delta_1=a_1a_3$ is $z$-independent, it is easy to decompose $1$ with respect to the dual basis:
	\begin{align}
		1 = \mu_1\Delta_1+\mu_2\Delta_2=(a_1a_3\mu_2)e_1+(a_1a_3\mu_1)e_2\,,
	\end{align}
	from which the dual basis coefficients are calculated to be \smash{$\mu_1=\frac{1}{a_1a_3}$ and $\mu_2=0$}.
\end{example}

\paragraph{Calculating Polynomial Inverses.}

Suppose we have some polynomial $q\in\mathbb{C}(\bm{a})[\bm{z}]$ which shares no common zeroes with $f_1,\ldots,f_n$. Introduce some auxiliary variable $w$ and define the ideal $\mathcal{J}=\langle f_1,\ldots,f_n, w q -1\rangle\subseteq \mathbb{C}(\bm{a})[w,z_1,\ldots,z_n]$. Let $\prec$ be the block order that compares the degrees of $w$ first and breaks ties using grlex or grevlex order on $\bm{z}$ with $z_1\succ\ldots\succ z_n$. Then inside $\mathcal{G}_{\prec}(\mathcal{J})$ there is a polynomial that is linear in $w$ of the form $w-q_\text{inv}(\bm{z})\in \mathcal{G}_{\prec}(\mathcal{J})$ where $q_\text{inv}$ is the polynomial inverse for $q$ in $Q$. The following example builds on \cref{ex:pf-formal}.

\begin{example}\label{ex:polynomial-inverses}
	Let $I$ be as in \cref{ex:pf-formal}. Suppose we wish to compute the polynomial inverse for $q(\bm{z})=z_1z_2$ in $Q$. Let $\mathcal{J}=\langle f_1,f_2,wq-1\rangle\subseteq\mathbb{C}(a_1,a_2,a_3)[w,z_1,z_2]$ and let $\prec$ be the block order described above. Then the Gr\"{o}bner basis $\mathcal{G}_\prec(\mathcal{J})$ for $\mathcal{J}$ with respect to $\prec$ contains the element \smash{$w+\frac{a_1a_3}{a_2}$} from which we can read of the polynomial inverse \smash{$q_\text{inv}(\bm{z})=-\frac{a_1a_3}{a_2}$}.
\end{example}

\paragraph{Push-Forwards.} Finally we return to the task of computing \eqref{eq:pf-coefficients-pushforward}. In particular, there are some simplifications which occur when using \eqref{eq:r-pushforward-gr} to compute  \eqref{eq:pf-coefficients-pushforward} for top-dimensional rational differential forms. To this end, consider \smash{$\underline{\omega}_{J}=\underline{\omega}^{\{1,\ldots,n\}}_{J}$}, for $J=\{j_1,\ldots,j_n\}$. Using \eqref{eq:r-pushforward-gr} together with  \eqref{eq:pf-coefficients} we find that
\begin{align}\label{eq:pf-coefficients-pushforward-residue}
	\mathcal{I}_\ast\underline{\omega}_{J} =\mathrm{Res}\bigg(\bigg|\frac{\partial\bm{f}}{\partial\bm{z}}\bigg|\,\underline{\omega}_{J}\bigg) =(-1)^n\,\mathrm{Res}\bigg(\underline{\omega}\,\bigg|\frac{\partial\bm{f}}{\partial\bm{a}}\bigg|_{J}\bigg)\,,
\end{align}
where $\underline{\omega}\in\mathbb{C}(\bm{z})$ is the rational function prefactor in \eqref{eq:omega} and \smash{$\big|{\partial\bm{f}}/{\partial\bm{a}}\big|_{J}$} is the maximal minor of the Jacobian matrix \smash{${\partial\bm{f}}/{\partial\bm{a}}$} specified by the columns $(j_1,\ldots,j_n)$. Notice that \smash{$\big|{\partial\bm{f}}/{\partial\bm{z}}\big|$} drops out of the global residue. Let \smash{$\underline{\omega}(\bm{z})={p(\bm{z})}/{q(\bm{z})}$} where $p$ and $q$ are polynomials such that $q$ does not vanish for any point in $\mathcal{V}(\mathcal{I})$, and let $q_\text{inv}$ be the polynomial inverse for $q$. Then combining \eqref{eq:pf-coefficients-pushforward-residue} with \eqref{eq:gr-final-r} we obtain the compact formula
\begin{align}\label{eq:pf-coefficients-pushforward-gr}
	\mathcal{I}_\ast\underline{\omega}_{J}= (-1)^n\,\mathrm{Res}\bigg(p\,q_\text{inv}\,\bigg|\frac{\partial\bm{f}}{\partial\bm{a}}\bigg|_{J}\bigg) = (-1)^n\sum_{\alpha=1}^{d}{\bigg(p\,q_\text{inv}\,\bigg|\frac{\partial\bm{f}}{\partial\bm{a}}\bigg|_{J}\bigg) }_\alpha\mu_\alpha\,.
\end{align}
This result is demonstrated in the following example.

\begin{example}\label{ex:pf-gr}
	Recall that in \cref{ex:pf-formal} we showed that
	the coefficients defined in \eqref{eq:pf-coefficients} and \eqref{eq:pf-coefficients-pushforward} are given by
	\begin{align*}
		\underline{\omega}_{\{1,2\}}=0\,,\qquad\underline{\omega}_{\{1,3\}}=\frac{1}{2 a_1 a_3}\,,\qquad \underline{\omega}_{\{2,3\}}=\frac{z_2}{2 a_1 a_3 z_1}\,,
	\end{align*}
	and
	\begin{align*}
		\mathcal{I}_\ast\underline{\omega}_{\{1,2\}}=0\,,\qquad \mathcal{I}_\ast\underline{\omega}_{\{1,3\}} = \frac{1}{a_1a_3}\,,\qquad \mathcal{I}_\ast\underline{\omega}_{\{2,3\}} = -\frac{1}{a_2a_3}\,,
	\end{align*}
	respectively.
	Let us re-compute these pushforward coefficients using \eqref{eq:pf-coefficients-pushforward-gr}. Recall from \cref{ex:pf-formal} that the Jacobian matrices are
	\begin{align}
		\frac{\partial\bm{f}}{\partial\bm{z}}=\begin{pmatrix}
			a_1 & a_2\\
			0 & 2 a_3z_2
		\end{pmatrix},\qquad \frac{\partial\bm{f}}{\partial\bm{a}}=\begin{pmatrix}
			z_1 & z_2 & 0\\
			0 & 0 & z_2^2
		\end{pmatrix}.
	\end{align}
 	Without loss of generality, assume grevlex-order where $z_1\succ z_2$.	In \cref{ex:dual-basis}, we found that \smash{$\mu_1=\frac{1}{a_1a_3}$} and $\mu_2=0$, while in \cref{ex:polynomial-inverses} we found that the polynomial inverse for $q(\bm{z})=z_1z_2$ in $Q$ is \smash{$q_\text{inv}(\bm{z})=-\frac{a_1a_3}{a_2}$}. Using \eqref{eq:pf-coefficients-pushforward-gr}, we have that
 \begin{align}
 	\mathcal{I}_\ast\underline{\omega}_{\{1,2\}} &= {\bigg(q_\text{inv}\,\bigg|\frac{\partial\bm{f}}{\partial\bm{a}}\bigg|_{\{1,2\}}\bigg)}_1\mu_1 = 0\,,\phantom{\frac 11}\\
 	\mathcal{I}_\ast\underline{\omega}_{\{1,3\}} &= {\bigg(q_\text{inv}\,\bigg|\frac{\partial\bm{f}}{\partial\bm{a}}\bigg|_{\{1,3\}}\bigg)}_1\mu_1 =-\frac{1}{a_2}{(z_1z_2^2)}_1 = \frac{1}{a_1a_3}\,,\\
 	\mathcal{I}_\ast\underline{\omega}_{\{2,3\}} &= \bigg(q_\text{inv}\,\bigg|\frac{\partial\bm{f}}{\partial\bm{a}}\bigg|_{\{2,3\}}\bigg)_1\mu_1 =-\frac{1}{a_2}(z_2^3)_1 = -\frac{1}{a_2a_3}\,,
 \end{align}
in agreement with the result above.

\end{example}

\section{Computational Considerations}\label{sec:computation}

Thus far we have presented three methods for computing the pushforward of a rational differential form through some zero-dimensional ideal $\mathcal{I}=\langle f_1,\ldots,f_n\rangle\subseteq\mathbb{C}(\bm{a})[\bm{z}]$ without needing to explicitly determine its common zeros $\mathcal{V}(\mathcal{I})$. In this section we compare the strengths and weaknesses of these methods from a computational point of view, and we give suggestions on how some of the bottlenecks can be overcome. To this end, let us briefly review the steps needed for each method to calculate the pushforward of a rational $p$-form $\omega = \underline{\omega}(\bm{z})\mathrm{d}z_{i_1}\wedge\cdots\wedge\mathrm{d}z_{i_p}$ through $\mathcal{I}$ where $\underline{\omega}\in\mathbb{C}(\bm{z})$. Recall that in \eqref{eq:pf-coefficients} we defined the rational functions
\begin{align*}
	\underline{\omega}^{I}_{J}(\bm{z};\bm{a})= (-1)^p\,\underline{\omega}(\bm{z})	\bigg|\bigg[\frac{\partial\bm{f}}{\partial\bm{z}}\bigg]^{-1}\frac{\partial\bm{f}}{\partial\bm{a}}\bigg|^{I}_{J}\in\mathbb{C}(\bm{a})(\bm{z})\,.
\end{align*}
The steps needed to calculate $\mathcal{I}_*\omega$ are as follows:
\begin{enumerate}[label=(\roman*)]
	\item \textbf{Companion Matrices:}\label{par:comp-matrices}
	\begin{enumerate}
		\item Find the Gr\"{o}bner basis $\mathcal{G}_\prec(\mathcal{I})$.
		\item Find the standard basis $\mathcal{B}_\prec(\mathcal{I})$.
		\item Find the companion matrices $T_i$.
		\item Evaluate the rational functions \smash{$\underline{\omega}^{I}_{J}$} on the companion matrices.
		\item Take the trace.
	\end{enumerate}
    % ----------
	\item \textbf{Derivatives of Companion Matrices:}\label{par:comp-matrices-derivatives}
	\begin{enumerate}
		\item Find the Gr\"{o}bner basis $\mathcal{G}_\prec(\mathcal{I})$. \label{itm:dcm-groebner}
		\item Find the standard basis $\mathcal{B}_\prec(\mathcal{I})$. \label{itm:dcm-standard-basis}
		\item Find the companion matrices $T_i$. \label{itm:dcm-comp}
		\item Calculate the partial derivatives of companion matrices with respect to $a$-variables.\label{itm:dcm-comp-der}
		\item Evaluate the rational function $\underline{\omega}$ on the companion matrices.\label{itm:dcm-rat-fn-on-cm}
		\item Evaluate \smash{$\sum_{\sigma\in S_p}\mathrm{sgn}(\sigma)\frac{\partial T_{i_{\sigma(1)}}}{\partial a_{j_1}}\cdots\frac{\partial T_{i_{\sigma(p)}}}{\partial a_{j_p}}$}. \label{itm:dcm-jac-factor-on-cm-derivatives}
		\item Multiply the results from \ref{itm:dcm-rat-fn-on-cm} and \ref{itm:dcm-jac-factor-on-cm-derivatives} and take the trace. \label{itm:dcm-trace}
	\end{enumerate}
	% ----------
	\item \textbf{Global Duality of Residues:}\label{par:global-residue}
	\begin{enumerate}
		\item Find the Gr\"{o}bner basis $\mathcal{G}_\prec(\mathcal{I})$.\label{itm:gr-groebner}
		\item Find the standard basis $\mathcal{B}_\prec(\mathcal{I})$.\label{itm:gr-standard-basos}
		\item Find the dual basis \smash{$\mathcal{B}^\vee_\prec(\mathcal{I})$} using the Bezoutian matrix.\label{itm:gr-dual-basis}
		\item Decompose unity in the dual basis.\label{itm:gr-decompose-unity}
		\item Calculate the polynomial inverses \smash{$\text{den}\big(\underline{\omega}^{I}_{J}\big)_{\text{inv}}$} of the denominators of \smash{$\underline{\omega}^{I}_{J}$}.\label{itm:gr-poly-inverses}
		\item Decompose
		\begin{align*}
			\overbar{\left|\frac{\partial\bm{f}}{\partial\bm{z}}\right|\text{den}\big(\underline{\omega}^{I}_{J}\big)_{\text{inv}}\,\text{num}\big(\underline{\omega}^{I}_{J}\big)}^{\mathcal{G}_\prec(\mathcal{I})},
		\end{align*}
		 in the standard basis.\label{itm:gr-decompose-fn}
		\item Take the dot product between the coefficients from \ref{itm:gr-decompose-unity} and \ref{itm:gr-decompose-fn}.\label{itm:gr-dot-prod}
	\end{enumerate}
\end{enumerate}

The first thing to note is that all methods require the calculation of a Gr\"obner basis $\mathcal{G}=\mathcal{G}_\prec(\mathcal{I})$ for $\mathcal{I}$ with respect to some monomial order $\prec$. At this point it is worth noting that the choice of monomial ordering $\prec$ can have significant influence on how long this takes. In practice, \emph{graded lexicographic} and \emph{graded reverse lexicographic} ordering are usually the best choice. For polynomial ideals with rational function coefficients, as in our case, the calculation of Gr\"obner bases can be an extremely costly operation, since most Gr\"{o}bner basis routines are optimised for implementations over finite fields. This becomes a significant bottleneck for large ideals, for example for the CHY scattering equations it already becomes near impossible to find an explicit Gr\"obner basis for $n>6$.

One possible way to significantly speed up the calculation of Gr\"obner bases is by using numerical values for the $a$-variables. The corresponding ideal $\mathcal{I}(\bm{a})$ is then simply an ideal in $\mathbb{C}[\bm{z}]$, and the computation of Gr\"obner basis can now be done much faster, using \emph{Faug\`ere's $F4$/$F5$ Algorithms} \cite{faugere1999new,faugere2002new}, for example. In this case, Gr\"obner bases for the scattering equations can easily be found up to $n=10$.

If we follow the above steps using numerical values for $a$-variables (making use of the methods described in section \cref{sec:alternative-companion-matrices} to calculate the derivatives of companion matrices), then the resulting rational prefactors \smash{$\mathcal{I}_*\underline{\omega}^{I}_{J}$} are numeric. There are a number of methods which can be used to reconstruct the algebraic answer from such numerical sampling. For example, if we set each $a_j$ to a distinct prime number, it is sometimes possible to reconstruct the algebraic answer using \emph{Egyptian fractions} from a single evaluation, see for example Section 4 of \cite{Huang:2015yka}. This does, however, only work if the rational prefactors are of a specific form. When this is not known to be the case, we can perform our numerical sampling over finite fields for which there exist multiple fast Gr\"{o}bner basis routines. In this case, the algebraic answer can be obtained using \emph{rational reconstruction} as implemented in \texttt{FiniteFlow} \cite{Peraro:2019svx} and \texttt{FireFly} \cite{Klappert:2019emp}. However, depending on the degree of the rational function and the number of variables, rational reconstruction may require many thousands of evaluations, which can take a significant amount of time.

In addition to these difficulties in calculating Gr\"{o}bner bases, there are several other points at which we can encounter computational bottlenecks. The three methods described above all have their own pros and cons, and therefore this is the point where one has to consider which of the methods is most appropriate for the task at hand.

The size of the companion matrices is $d\times d$ where $d=|\mathcal{V}(\mathcal{I})|=|\mathcal{B}|=\dim(Q)$. Depending on the size of the variety, the companion matrices can become very large. For example, the companion matrices for the $n$-particle CHY scattering equations are of size $(n-3)!\times(n-3)!$. There is a potential bottleneck in evaluating a rational function on big companion matrices, depending on the complexity of the rational function. This is the place where the main difference between methods \ref{par:comp-matrices} and \ref{par:comp-matrices-derivatives} is revealed, since \ref{par:comp-matrices-derivatives} only has a single rational function to evaluate whereas \ref{par:comp-matrices} will in general have a large list of functions to evaluate. Furthermore, the rational functions one has to evaluate on the companion matrices for methods  \ref{par:comp-matrices} and \ref{par:comp-matrices-derivatives} differs by a factor of \smash{$\big|\big[\frac{\partial\bm{f}}{\partial\bm{z}}\big]^{-1}\frac{\partial\bm{f}}{\partial\bm{a}}\big|^{I}_{J}$}. This can in general be a very unwieldy expression, and can significantly impact the calculation time. These considerations favour method \ref{par:comp-matrices-derivatives} over \ref{par:comp-matrices}, but the method \ref{par:comp-matrices-derivatives} has some additional drawbacks. Specifically, it requires us to find the derivatives of companion matrices, and to evaluate complicated product expressions for them as in point \ref{itm:dcm-jac-factor-on-cm-derivatives}. When doing calculations on numerical data, we have to use the method described in \cref{sec:alternative-companion-matrices} to find the derivatives of companion matrices. This requires additional Gr\"obner basis calculations in a larger ideal, and, importantly, they have to be done using \emph{lexicographic} monomial ordering, which takes significantly longer.

These issues are sidestepped by method \ref{par:global-residue}, since it does not require us to evaluate a rational function on large matrices. It does, however, have its own set of limitations. The first difficulty we can encounter is in finding the dual basis $\mathcal{B}^\vee$. As explained in \cref{sec:global-residue}, to calculate the dual basis we first have to find the Bezoutian matrix $B$, which is of size $n \times n$ (where $n$ indicates the number of functions that generate our ideal). Finding $B$ and its determinant $\det{B}$ is not too difficult, however finding the remainder of $\det{B}$ with respect to $\mathcal{G}\cup \tilde{\mathcal{G}}$ can be a time intensive task. The second drawback of this method is in calculating the polynomial inverse \smash{$\text{den}\big(\underline{\omega}^{I}_{J})_{\text{inv}}$}. This again requires an additional Gr\"obner basis calculation of a slightly larger ideal, which, as stressed before, can be quite time intensive. One simplification occurs when calculating the pushforward of a top-form:
\begin{align}
	(-1)^p\,\underline{\omega}(\bm{z})\left|\frac{\partial \bm{f}}{\partial{\bm{z}}}\right|\,	\bigg|\bigg[\frac{\partial\bm{f}}{\partial\bm{z}}\bigg]^{-1}\frac{\partial\bm{f}}{\partial\bm{a}}\bigg|_{J} = (-1)^p\,\underline{\omega}\left|\frac{\partial\bm{f}}{\partial\bm{a}}\right|_{J},
\end{align}
and the resulting rational function whose remainder we have to decompose in $\mathcal{B}$ is much simpler. Now only the polynomial inverse of the denominator of $\underline{\omega}$ needs to be calculated. Indeed, if $\omega$ is a polynomial top-form, then no polynomial inverses have to be calculated at all!

\section{Examples}\label{sec:examples}

\subsection{ABHY Associahedron and Bi-adjoint \texorpdfstring{$\phi^3$}{Cubic Scalar Theory} Amplitudes}\label{sec:examples-ABHY}

In this section we will use the methods developed in this paper to derive formulae for the \emph{ABHY associahedron} canonical form from which we can subsequently find formulae for \emph{bi-adjoint $\phi^3$ amplitudes}.

The starting point for our discussion are the original \emph{Cachazo--He--Yuan (CHY) scattering equations} \cite{Cachazo:2013gna}. They are a universal system of equations which connect the moduli space of the $n$-punctured Riemann sphere to the kinematic space $\mathbb{K}_n$ which governs the scattering of $n$ massless particles. The moduli space is parametrized by the positions of  $n$-punctures $z_1,\ldots,z_n$ modulo an $\mathsf{SL}(2,\mathbb{C})$ symmetry. The kinematic space $\mathbb{K}_n$ is spanned by linearly independent Mandelstam variables in $d\ge n-1$ dimensions. Given any set of particle labels $A\in\binom{[n]}{\ell}$, where $\ell=1,\ldots,n-3$, the Mandelstam variable $s_A$ is defined by
\begin{align}
	s_A \coloneqq \Big(\sum_{i\in A}k_i\Big)^2 = \sum_{\mathclap{i<j\in A}} s_{ij}\,,\qquad s_{ij}\coloneqq (k_i+k_j)^2 = 2k_i\cdot k_j\,,
\end{align}
where $k_i^\mu$ are $n$ massless momenta satisfying momentum conservation $\sum_{i=1}^nk_i^\mu=0$. The dimensionality of $\mathbb{K}_n$ is \cite{Arkani-Hamed:2017mur}
\begin{align}
	\dim\mathbb{K}_n = \binom{n}{2} -n = \frac{n(n-3)}{2}\,.
\end{align}
Of special interest to us are the planar Mandelstam variables which are defined by
\begin{align}
	X_{i,j}\coloneqq s_{i,i+1,\ldots,j-1}\,,
\end{align}
where $1\le i<j\le n$ and which vanish identically when $j=i+1$ or $(i,j)=(1,n)$. Let us denote the set of non-vanishing planar Mandelstam variables by $\mathcal{X}_n$ of which there are precisely ${n(n-3)}/{2}$. Then $\mathcal{X}_n$ furnishes a natural basis for $\mathbb{K}_n$ since each two-particle Mandelstam variable can be expanded in terms of $\mathcal{X}_n$ via
\begin{align}
	s_{i,j}= X_{i,j+1}+X_{i+1,j}-X_{i,j} - X_{i+1,j+1}\,.
\end{align}
We will sometimes use the notation $a_j$ to denote the $j$\textsuperscript{th} non-vanishing planar Mandelstam variable
where
\begin{align}
	X_{i,j} = a_{(i-1)n+j+\delta_{i,1}-\binom{i+2}{2}}\,,
\end{align}
and \smash{$\bm{a}=(a_1,\ldots,a_{|\mathcal{X}_n|})$}. In \cite{Dolan:2014ega}, Dolan and Goddard showed that the scattering equations can be recast in polynomial form using the polynomials
\begin{align}
	h_\ell \coloneqq \sum_{A\in\binom{[n]}{\ell+1}} s_A z_A\,,
\end{align}
where $z_A\coloneqq \prod_{i\in A}z_i$ for $\ell=1,\ldots,n-3$. Throughout this section, we will use the $\mathsf{SL}(2,\mathbb{C})$ symmetry of the scattering equations to fix $z_1\to0,z_{n-1}\to1,z_n\to\infty$, in which case we obtain
\begin{align}
	f_\ell\coloneqq \lim_{z_n\to\infty} \left.\frac{h_\ell}{z_n}\right|_{z_1\to 0\,,\,z_{n-1}\to 1}.
\end{align}
In terms of these gauge-fixed polynomials, the scattering equations define the following ideal
\begin{align}
	\mathcal{I}_n\coloneqq\langle f_1,\ldots,f_{n-3}\rangle\subseteq\mathbb{Q}(\mathcal{X}_n)[\bm{z}]\,,
\end{align}
where $\bm{z}=(z_2,\ldots,z_{n-2})$.

For the above choice of gauge fixing, the \emph{$n$-particle world-sheet Parke-Taylor form} is the $(n-3)$-form
\begin{align}
	\omega^{\text{WS}}_n\coloneqq- \frac{\mathrm{d}z_2\wedge\cdots\wedge\mathrm{d}z_{n-2}}{z_2(z_2-z_3)\cdots(z_{n-2}-1)}.
\end{align}
It was shown in \cite{Arkani-Hamed:2017mur} that the canonical form of the so-called ABHY associahedron can be expressed as the pushforward of $\omega^{\text{WS}}_n$ through the scattering equations:
\begin{align}\label{eq:ABHY-pushforward}
	\omega^{\text{ABHY}}_n = \mathcal{I}_{n\ast} \omega^{\text{WS}}_n.
\end{align}
It was further shown in \cite{Arkani-Hamed:2017mur} that the $n$-point bi-adjoint $\phi^3$ amplitude can be extracted from $\omega^{\text{ABHY}}_n$ by pulling it back to an $(n-3)$-dimensional subspace defined by the relations
\begin{align}
c_{i,j} = X_{i,j}+X_{i+1,j+1}-X_{i+1,j}-X_{i,j+1}\,,
\end{align}
for every non-adjacent pair of indices $1\le i < j \le n-1$ where $c_{i,j}$ are positive constants. These relations can be used to solve for $X_{i>1,j}$ in terms of $X_{1,3},\ldots,X_{1,n-1}$:
\begin{align}
	X_{i>1,j} = X_{1,j}-X_{1,i+1} + C_{i,j},
\end{align}
where $C_{i,j}\coloneqq\sum_{r=1}^{i-1}\sum_{s=i+1}^{j-1} c_{r,s}$ is a positive constant. Let us use $b_i=X_{1,i}$ to enumerate the coordinates which parametrise the subspace onto which we pull back. The $n$-point scattering amplitude $m_n$ can then be calculated as \cite{Arkani-Hamed:2017mur}
\begin{align}\label{eq:ABHY-pullback}
	\mathcal{J}_n^\ast\omega^{\text{ABHY}}_n = m_n\bigwedge_{k=3}^{n-1} \mathrm{d}b_k\,,
\end{align}
where the ideal through which we pull back is given by
\begin{align}
\mathcal{J}_n=\left\langle
X_{i,j} - g_{i,j}(\bm{b}):X_{i,j}\in\mathcal{X}_n
\right\rangle
\subseteq\mathbb{Q}(\bm{b})[\bm{z}]\,,
\end{align}
with $\bm{b}=(b_3,\ldots,b_{n-1})$,
$g_{1,i}(\bm{b})=b_i$ and $g_{i>1,j}(\bm{b})= b_{j}-b_{i+1}+C_{i,j}$.

Using the notation of \eqref{eq:pf-form-with-coefficients}, we can express the pushforward in \eqref{eq:ABHY-pushforward} as
\begin{align} \label{eq:ABHY-pushforward-with-coefficients}
	\omega^{\text{ABHY}}_n = \mathcal{I}_{n\ast} \omega^{\text{WS}}_n = \sum_{J\in \binom{[|\mathcal{X}_n|]}{n-3}} \mathcal{I}_{n\ast} \underline{\omega}_{J} \bigwedge_{j\in J}\mathrm{d}a_{j}\,,
\end{align}
where the coefficients $\underline{\omega}_{J}(\bm{z})$ are calculated according to \eqref{eq:pf-coefficients} as
\begin{align}
	\underline{\omega}_{J}(\bm{z}) = \frac{(-1)^{n}}{z_2(z_2-z_3)\cdots(z_{n-2}-1)} \left|\frac{\partial\bm{f}}{\partial\bm{z}}\right|^{-1}\left|\frac{\partial\bm{f}}{\partial\bm{a}}\right|_{J}.
\end{align}
The pushforward can then be evaluated using the methods highlighted in \cref{sec:comp-matrices,sec:global-residue} to sum $\underline{\omega}_{J}(\bm{z})$ over $\mathcal{V}(\mathcal{I}_n)$. We can also combine \eqref{eq:ABHY-pullback} with \eqref{eq:ABHY-pushforward} to obtain the following formula for $m_n$:
\begin{align}\label{eq:ABHY-pf-of-pb-of-SE}
	m_n\bigwedge_{\mathclap{k=3}}^{\mathclap{n-1}} \mathrm{d}b_k = \mathcal{J}_n^\ast\omega^{\text{ABHY}}_n = \mathcal{J}_n^\ast(\mathcal{I}_{n\ast} \omega^{\text{WS}}_n)
	= (\mathcal{J}_n^\ast\mathcal{I}_n)_\ast A_{n}\bigwedge_{\mathclap{k=3}}^{\mathclap{n-1}} \mathrm{d}b_k\,,
\end{align}
where
\begin{align}\label{eq:ABHY-A}
	\begin{split}
A_{n}(\bm{z};\bm{b})\coloneqq& \sum_{J\in \binom{[|\mathcal{X}_n|]}{n-3}}  \mathcal{J}_n^\ast\underline{\omega}_{J}(\bm{z})\left|\frac{\partial\bm{g}}{\partial\bm{b}}\right|^{J}\\
=& \frac{(-1)^{n}}{z_2(z_2-z_3)\cdots(z_{n-2}-1)} \left|\frac{\partial(\mathcal{J}_n^\ast\bm{f})}{\partial\bm{z}} \right|^{-1}\sum_{J\in \binom{[|\mathcal{X}_n|]}{n-3}} \left|\frac{\partial\bm{f}}{\partial\bm{a}}\right|_{J}\left|\frac{\partial\bm{g}}{\partial\bm{b}}\right|^{J}\\
=&\frac{(-1)^{n}}{z_2(z_2-z_3)\cdots(z_{n-2}-1)} \left|\frac{\partial(\mathcal{J}_n^\ast\bm{f})}{\partial\bm{z}} \right|^{-1}\left|\frac{\partial (\mathcal{J}_n^*\bm{f})}{\partial \bm{b}}\right|,
	\end{split}
\end{align}
is a rational function in $\mathbb{Q}(\bm{b})(\bm{z})$,
\begin{align}
	\mathcal{J}_n^\ast f_\ell = \left.f_\ell\right|_{X_{i,j}\to g_{i,j}(\bm{b})},
\end{align}
is the pullback of the polynomial $f_\ell$ through $\mathcal{J}_n$, and
\begin{align}
	\mathcal{J}_n^\ast\mathcal{I}_n \coloneqq \left\langle\mathcal{J}_n^\ast f_1,\ldots,\mathcal{J}_n^\ast f_{n-3}\right\rangle\subseteq\mathbb{Q}(\bm{b})[\bm{z}]\,,
\end{align}
is the ideal generated by the polynomials $\mathcal{J}_n^\ast f_1,\ldots,\mathcal{J}_n^\ast f_{n-3}$. The last equality in \eqref{eq:ABHY-A} follows from the simplification
\begin{align}
\sum_{J\in \binom{[|\mathcal{X}_n|]}{n-3}} \left|\frac{\partial(\mathcal{J}_n^\ast\bm{f})}{\partial\bm{a}}\right|_{J}\left|\frac{\partial\bm{g}}{\partial\bm{b}}\right|^{J} = \left|\frac{\partial (\mathcal{J}_n^*\bm{f})}{\partial \bm{b}}\right|.
\end{align}
The final expression in \eqref{eq:ABHY-A} essentially means that, at least in this case, the pullback and pushforward are `associative' in the sense that $\mathcal{J}^*(\mathcal{I}_*\,\omega) = (\mathcal{J}^*\mathcal{I})_*\omega$.
From the polynomial scattering equations one can show that
\begin{align}\label{eq:ABHY-df/db}
	\left|\frac{\partial(\mathcal{J}_n^*\bm{f})}{\partial\bm{b}}\right| = (-1)^{n+1} \prod_{i=1 }^{n-3}\prod_{j= i+2 }^{n-1}(z_i-z_j)\,,
\end{align}
with $z_1=0$ and $z_{n-1}=1$, which is independent of $\bm{b}$. To find the coefficient $m_n$ of $\mathrm{d}b_3\wedge\cdots\wedge\mathrm{d}b_{n-1}$ in \eqref{eq:ABHY-pf-of-pb-of-SE} one has to sum $A_n$ over $\mathcal{V}(\mathcal{J}_n^\ast\mathcal{I}_n)$
\begin{align}
	m_n = \sum_{\bm{\xi} \in \mathcal{V}(\mathcal{J}_n^\ast\mathcal{I}_n)} A_n(\bm{\xi})\,,
\end{align}
which gives a rational function in $\bm{b}$ and $C_{i,j}$. We can then re-express $m_n$ as a rational function in $X_{i,j}$ by re-substituting $b_i\to X_{1,i}$ and $C_{i,j}\to X_{i,j}+X_{1,i+1}-X_{1,j}$. The latter result can also be achieved by replacing $|\partial(\mathcal{J}_n^*\bm{f})/\partial\bm{z}|$ with $ |\partial\bm{f}/\partial\bm{z}|$ in $A_n$ and instead summing over $\mathcal{V}(\mathcal{I}_n)$ to produce the following formula for $m_n$:
\begin{align}\label{eq:ABHY-phi-cubed-amplitude}
	m_n=\sum_{\bm{\xi} \in \mathcal{V}(I_n)} A'_n(\bm{\xi}),
\end{align}
where
\begin{align}\label{eq:ABHY-amplitude-summand}
	A'_n(\bm{z})=-\left|\frac{\partial\bm{f}}{\partial\bm{z}}\right|^{-1}\frac{\prod_{i=1 }^{n-3}\prod_{j= i+2 }^{n-1}(z_i-z_j)}{z_2(z_2-z_3)\cdots(z_{n-2}-1)}\,,
\end{align}
is a rational function in $\mathbb{Q}(\mathcal{X}_n)(\bm{z})$.

We conclude this part by noticing that the formula \eqref{eq:ABHY-phi-cubed-amplitude} looks remarkably similar to the CHY formula for the $n$-point amplitude scattering amplitudes in bi-adjoint $\phi^3$ theory from \cite{Cachazo2014}. As we will show shortly, the two formulae are in fact identical, and their apparent differences come from the different formulations of the scattering equations. Indeed, we can find a new CHY-like formula for bi-adjoint $\phi^3$ amplitudes for each equivalent form of the scattering equations. Specifically, if the functions $h_1,\ldots,h_{n-3}$ have the same common zeroes as the scattering equations, then using the associativity of the pullback and pushforward, we can find the following CHY summand:
\begin{align}\label{eq:conclusion-summand}
	A_n(\bm{z})\coloneqq \frac{(-1)^n}{(z_1-z_2)(z_2-z_3)\cdots(z_n-z_1)}\left|\frac{\partial\bm{h}}{\partial \bm{z}}\right|^{-1}\left|\frac{\partial(\mathcal{J}_n^*\bm{h})}{\partial \bm{b}}\right|/\mathsf{SL}(2)_z,
\end{align}
where the $\mathsf{SL}(2)_z$ instructs us to fix the position of three of the $z$'s. The amplitude is then found as
\begin{align}
	m_n=\sum_{\bm{\xi}\in\mathcal{V}(\mathcal{I}_n)}A_n(\bm{\xi})\,,
\end{align}
where $\mathcal{I}_n$ is now the ideal generated by the polynomial scattering equations with the appropriate gauge fixing.
To make the connection to the standard CHY formula, let us define
\begin{align}
	\tilde{h}_a\coloneqq \sum_{b\neq a} \frac{s_{ab}}{z_a-z_b},\quad a=1,\ldots,n\,,
\end{align}
to be the original CHY scattering equations \cite{Cachazo:2013gna}. Only $n-3$ of these equations are independent, and we can thus remove any 3 of these equations, say $\tilde{h}_q,\,\tilde{h}_r$, and $\tilde{h}_s$. Explicit calculation yields
\begin{align}
	\left|\frac{\partial(\mathcal{J}_n^* \bm{\tilde{h}})}{\partial\bm{b}}\right|^{[n]\setminus\{q,r,s\}} = (-1)^{n+q+r+s+1}\frac{(z_q-z_r)(z_r-z_s)(z_s-z_q)}{(z_1-z_2)(z_2-z_3)\cdots(z_n-z_1)},
\end{align}
and equation \eqref{eq:conclusion-summand} turns into the standard CHY summand. We thus see that from this point of view, the fundamental quantity in the CHY summand is only a single Parke-Taylor factor multiplied by Jacobians that depend on the specific form of the scattering equations, which may or may not yield a second Parke-Taylor form. As noted before, for polynomials $h_a$ we can apply the method of global residues (\cref{sec:global-residue}) using these equations, in which case the Jacobian $\smash{|\partial\bm{h}/\partial\bm{z}|^{-1}}$ drops out of the calculation. It would be interesting to see if one can use the results of this paper to derive similar formulae for other CHY-constructable theories. 

To illustrate the results from this section, we will explicitly compute $\omega_n^\text{ABHY}$ and $m_n$ for $n=4$ and $n=5$.

\paragraph{Four-points.}
The $n=4$ scattering equations are expressed via a single polynomial
$f_1=X_{1,3}-(X_{1,3}+X_{2,4})z_2$
which defines the ideal $\mathcal{I}_4=\langle f_1\rangle\subseteq \mathbb{Q}(\mathcal{X}_4)[z_2]$
where $\mathcal{X}_4=\{X_{1,3},X_{2,4}\}$. In this case, the associated variety is trivially found to be
\begin{align}\label{eq:ABHY-4-variety}
\mathcal{V}(\mathcal{I}_4)=\left\{\frac{X_{1,3}}{X_{1,3}+X_{1,4}}\right\}.
\end{align}
We are interested in calculating the pushforward of
\begin{align}
\omega_4^\text{WS}=-\frac{\mathrm{d}z_2}{z_2(z_2-1)}\,,
\end{align}
through $\mathcal{I}_4$ which can be expressed as
\begin{align}
\omega_4^\text{ABHY} = \mathcal{I}_{4\ast}\omega_4^\text{WS} = (\mathcal{I}_{4\ast}\underline{\omega}_{\{1\}})\,\mathrm{d}X_{1,3}+ (\mathcal{I}_{4\ast}\underline{\omega}_{\{2\}})\,\mathrm{d}X_{2,4}\,,
\end{align}
where
\begin{align}
\underline{\omega}_{\{1\}}(z_2) = \frac{1}{z_2(X_{1,3}+X_{2,4})}\,,\qquad\underline{\omega}_{\{2\}}(z_2) = \frac{1}{(z_2-1)(X_{1,3}+X_{2,4})}\,.
\end{align}
Here the subscripts $\{1\}$ and $\{2\}$ refer to the labelling of the planar Mandelstam variables $(a_1,a_2)=(X_{1,3},X_{2,4})$. Using \eqref{eq:ABHY-4-variety}, we find by direct computation that
\begin{align}
	\mathcal{I}_{4\ast}\underline{\omega}_{\{1\}} = \frac{1}{X_{1,3}}\,,\qquad\mathcal{I}_{4\ast}\underline{\omega}_{\{2\}} = -\frac{1}{X_{2,4}}\,,
\end{align}
and hence
\begin{align}\label{eq:ABHY-n=4-form}
\omega_4^{\text{ABHY}}=\frac{\mathrm{d}X_{1,3}}{X_{1,3}}-\frac{\mathrm{d}X_{2,4}}{X_{2,4}}\,,
\end{align}
in agreement with the known result. To extract the four-point scattering amplitude $m_4$, we pull $\omega_4^{\text{ABHY}}$ back through the ideal
$\mathcal{J}_4=\langle X_{1,3}-g_{1,3}(b_3), X_{2,4}-g_{2,4}(b_3) \rangle\subseteq\mathbb{Q}(b_3)[\mathcal{X}_4]$
where $b_3=X_{1,3}$, $g_{1,3}(b_3)=b_3$ and $g_{2,4}(b_3) = C_{2,4}-b_3$, to obtain
\begin{align}
	\mathcal{J}_4^* \omega_4^{\text{ABHY}} = m_4 \mathrm{d}b_3\,.
\end{align}
Consequently
\begin{align}\label{eq:phi-cubed-4-point-amp}
m_4= \mathcal{J}_4^\ast(\mathcal{I}_{4\ast}\underline{\omega}_{\{1\}})\left|\frac{\partial\bm{g}}{\partial b_3}\right|_{\{1\}}+ \mathcal{J}_4^\ast(\mathcal{I}_{4\ast}\underline{\omega}_{\{2\}})\left|\frac{\partial\bm{g}}{\partial b_3}\right|_{\{2\}} = \frac{1}{b_3}+\frac{1}{C_{2,4}-b_3}\,.
\end{align}
Re-substituting $b_3\to X_{1,3}$ and $C_{2,4}\to X_{1,3}+X_{2,4}$ we then recognise the four-point scattering amplitude
\begin{align}
m_4=\frac{1}{X_{1,3}}+\frac{1}{X_{2,4}}\,.
\end{align}

\paragraph{Five-points.}

The above example was trivial since the variety was one-dimensional and easy to determine by inspection. The case when $n=5$ is comparatively more involved and we will calculate $\omega_5^{\text{ABHY}}$ and $m_5$ using the method introduced in \cref{sec:global-residue} which leverages the global duality of residues.

The $n=5$ scattering equations are expressed via the polynomials
\begin{align}
	\begin{split}
	f_1&=-(X_{1,3}+X_{2,5}-X_{3,5})z_2+(X_{1,3}-X_{1,4}-X_{3,5})z_3+X_{1,4}\,,\\
	f_2&=-(X_{1,4}-X_{2,4}+X_{2,5})z_2 z_3-(X_{1,3}-X_{1,4}+X_{2,4})z_2+X_{1,3}z_3\,,
	\end{split}
\end{align}
where we will sometimes label $(a_1,a_2,a_3,a_4,a_5)=(X_{1,3},X_{1,4},X_{2,4},X_{2,5},X_{3,5})$. These polynomials generate the ideal $\mathcal{I}_5=\langle f_1, f_2 \rangle\subseteq \mathbb{Q}(\mathcal{X}_5)[\bm{z}]$ where $\bm{z}=(z_2,z_3)$. By pushing
\begin{align}
	\omega_5^{\text{WS}}=-\frac{\mathrm{d}z_2\wedge\mathrm{d}z_3}{z_2(z_2-z_3)(z_3-1)}\,,
\end{align}
forward through $\mathcal{I}_5$ we obtain
\begin{align}
	\omega_5^{\text{ABHY}}= \mathcal{I}_{5\ast} \omega_5^{\text{WS}} = \sum_{\mathclap{J=\{j_1,j_2\}\in\binom{[5]}{2}}}\,\mathcal{I}_{5\ast} \underline{\omega}_{\{j_1,j_2\}}\,\mathrm{d}a_{j_1}\wedge\mathrm{d}a_{j_2}\,.
\end{align}
where $\underline{\omega}_{J}$ are calculated according to \eqref{eq:pf-coefficients} as
\begin{align}
\underline{\omega}_{J}(\bm{z}) = - \frac{1}{z_2(z_2-z_3)(z_3-1)}\left|\frac{\partial\bm{f}}{\partial\bm{z}}\right|^{-1} \left|\frac{\partial\bm{f}}{\partial\bm{a}}\right|_{J}.
\end{align}
Explicitly,
\begin{align}\label{eq:ABHY-n=5-omega_ij}
\underline{\omega}_{J}(\bm{z})
=\left(\frac{1-z_2}{z_2},1,-1,\frac{z_2-z_3}{z_2(1-z_3)},\frac{1-z_3}{z_3-z_2},1,-1,\frac{z_2}{z_3-z_2},1,\frac{z_3}{1-z_3}\right)\left|\frac{\partial\bm{f}}{\partial\bm{z}}\right|^{-1},
\end{align}
where
\begin{align}\label{eq:ABHY-n=5-df/dz}
	\smash{\left|\frac{\partial\bm{f}}{\partial\bm{z}}\right|} =&
	(X_{1,4}-X_{2,4}) (X_{1,4}+X_{3,5})-X_{1,3} (2 X_{1,4}-X_{2,4}+X_{2,5})\\
&+ (X_{1,4}-X_{2,4}+X_{2,5})\left(z_2X_{2,5}+(z_2+z_3) (X_{1,3}-X_{3,5})-z_3X_{1,4}\right).\nonumber
\end{align}
We will sum $\underline{\omega}_J$ functions over all points in $\mathcal{V}(\mathcal{I}_5)$ using the methods introduced in \cref{sec:global-residue}. For the readers convenience, we will explicitly follow the steps outlined in \cref{sec:computation}.
\begin{enumerate}
	\item[\ref{itm:gr-groebner}] Choosing grevlex ordering with $z_2\succ z_3$, we calculate the Gr\"obner basis to be
	\begin{align}
	&\mathcal{G}\coloneqq\mathcal{G}_{\text{grevlex}}(\mathcal{I}_5)=\\
	\nonumber&\left\{
	\begin{array}{l}
		z_2+
		\frac{(z_3-1) X_{1,4}+z_3 (X_{3,5}-X_{1,3})}{X_{1,3}+X_{2,5}-X_{3,5}}\,, \\
		z_3^2-2z_3+\frac{z_3 X_{2,5}}{X_{1,4}-X_{2,4}+X_{2,5}}-\frac{z_3 X_{3,5}}{X_{1,3}-X_{1,4}-X_{3,5}}-\frac{z_3 X_{1,3} X_{2,4}-X_{1,4} (X_{1,3}-X_{1,4}+X_{2,4})}{(X_{1,4}-X_{2,4}+X_{2,5}) (X_{1,3}-X_{1,4}-X_{3,5})}
	\end{array}
	\right\}.
	\end{align}
	\item[\ref{itm:gr-standard-basos}] The leading monomials of polynomials in the Gr\"obner basis are $\{z_3^2,z_2\}$, and hence the standard basis of $Q\coloneqq\mathbb{Q}(\mathcal{X}_5)[\bm{z}]/\mathcal{I}_5$ is given by $\mathcal{B}\coloneqq\mathcal{B}_{\text{grevlex}}(\mathcal{I}_5)=\{e_1,e_2\}\coloneqq\{z_3,1\}$.
	\item[\ref{itm:gr-dual-basis}] The dual basis is given by $\mathcal{B}^\vee\coloneqq \mathcal{B}^\vee_{\text{grevlex}}(\mathcal{I}_5)=\{\Delta_1,\Delta_2\}$ where
	\begin{subequations}\label{eq:ABHY-n=5-dual-basis}
	\begin{align}
		\Delta_1&=(X_{1,4}-X_{2,4}+X_{2,5})(X_{1,3}-X_{1,4}-X_{3,5})\,,\\
		\Delta_2&= (z_3-1) \Delta_1-X_{1,4} (X_{1,3}-X_{1,4}+X_{2,4})-X_{2,5} X_{3,5}\,,
	\end{align}
	\end{subequations}
	are calculated using \eqref{eq:bez-dual-basis}.
	\item[\ref{itm:gr-decompose-unity}] Since $\Delta_1$ is constant with respect to $\bm{z}$, the coefficients of the decomposition of unity in terms of the dual basis $\mathcal{B}^\vee$ is given by $1=\mu_1\Delta_1+\mu_2\Delta_2$ where $\mu_1=1/\Delta_1$ and $\mu_2=0$.
	\item[\ref{itm:gr-poly-inverses}] Having determined $\mathcal{G}$, $\mathcal{B}$, $\mathcal{B}^\vee$ and $\Delta_{1,2}$, we now need to compute the polynomial inverses of the denominators
	\begin{align}
	\underline{\omega}_{J} \left|\frac{\partial \bm{f}}{\partial \bm{z}}\right| = - \frac{1}{z_2(z_2-z_3)(z_3-1)} \left|\frac{\partial\bm{f}}{\partial\bm{a}}\right|_{J}\,,
	\end{align}
	where the Jacobian \smash{$\big|\frac{\partial \bm{f}}{\partial \bm{z}}\big|$} cancels the inverse Jacobian present in $\underline{\omega}_{J}$ as seen in \eqref{eq:ABHY-n=5-omega_ij}. There are only three distinct denominators: $z_2$, $1-z_3$ and $z_3-z_2$. Their polynomial inverses are evaluated using the procedure described in \cref{sec:global-residue} and are given by
	\begin{subequations}
		\label{eq:ABHY-n=5-poly-inverses}
	\begin{alignat}{2}
		&(z_2)_\text{inv}&=&\textstyle\frac{X_{1,3} (2 X_{1,4}-X_{2,4}+X_{2,5})+X_{1,4} (X_{2,5}-X_{3,5})+X_{2,4} X_{3,5}-z_2\Delta'}{X_{1,3} X_{1,4}}\,,\\
		&(1-z_3)_\text{inv} &=&\textstyle\frac{X_{1,3} X_{1,4}+X_{1,4} X_{2,5}+X_{2,5} X_{3,5}-z_2\Delta'}{X_{2,5} X_{3,5}}\,,\\
		&(z_3-z_2)_\text{inv} &=&\textstyle\frac{\left(X_{1,4}+X_{2,5}\right) \left(X_{1,3} (X_{1,4}+X_{2,5})-z_2\Delta'\right)-X_{2,4} X_{2,5} (X_{1,3}-X_{1,4}-X_{3,5})}{X_{1,4} X_{2,4} X_{2,5}}\,.
	\end{alignat}
\end{subequations}
	where
	\begin{align}
		\Delta' = (X_{1,3}+X_{2,5}-X_{3,5}) (X_{1,4}-X_{2,4}+X_{2,5})\,.
	\end{align}
	\item[\ref{itm:gr-decompose-fn}] Since $\mu_2=0$, we only need the coefficient of \smash{$\underline{\omega}_J\big|{\partial \bm{f}}/{\partial \bm{z}}\big|$} (with denominator factors replaced with their polynomial inverses) modulo $\mathcal{G}$ multiplying $e_1=z_3$. In particular, we find that
	\begin{subequations}
		\label{eq:ABHY-n=5-fn-decomp}
		\allowdisplaybreaks
	\begin{alignat}{3}
	&\{1,2\}:&&\left(\overbar{(1-z_2)(z_2)_{\text{inv}}}^{\mathcal{G}}\right)_1&=&-\frac{\Delta_1}{X_{1,3} X_{1,4}}\,,\\
	&\{1,5\}:&&\left(\overbar{(z_2-z_3)(z_2)_{\text{inv}}(1-z_3)_{\text{inv}}}^{\mathcal{G}}\right)_1&=&+\frac{\Delta_1}{X_{1,3} X_{3,5}}\,,\\
	&\{2,3\}:&&\left(\overbar{(1-z_3)(z_3-z_2)_{\text{inv}}}^{\mathcal{G}}\right)_1&=&-\frac{\Delta_1}{X_{1,4} X_{2,4}}\,,\\
	&\{3,4\}:&&\left(\overbar{z_2(z_3-z_2)_{\text{inv}}}^{\mathcal{G}}\right)_1&=&-\frac{\Delta_1}{X_{2,4} X_{2,5}}\,,\\
	&\{4,5\}:&&\left(\overbar{z_3(1-z_3)_{\text{inv}}}^{\mathcal{G}}\right)_1&=&-\frac{\Delta_1}{X_{2,5} X_{3,5}}\,,
	\end{alignat}
\end{subequations}
	while the relevant coefficients for non-cyclic subsets $J\in\binom{[5]}{2}$ vanish since \smash{$\big(\overbar{\pm1}^{\mathcal{G}}\big)_1=0$}.
	\item[\ref{itm:gr-dot-prod}] Combining everything together, multiplying the expressions in \eqref{eq:ABHY-n=5-fn-decomp} by $\mu_1=\frac{1}{\Delta_1}$, we finally obtain
	\begin{align}\label{eq:ABHY-n=5-form}
		&\omega_5^{\text{ABHY}}= \\
		\nonumber&-\frac{\mathrm{d}X_{1,3}\wedge \mathrm{d}X_{1,4}}{X_{1,3} X_{1,4}}+\frac{\mathrm{d}X_{1,3}\wedge \mathrm{d}X_{3,5}}{X_{1,3} X_{3,5}}-\frac{\mathrm{d}X_{1,4}\wedge \mathrm{d}X_{2,4}}{X_{1,4} X_{2,4}}-\frac{\mathrm{d}X_{2,4}\wedge \mathrm{d}X_{2,5}}{X_{2,4} X_{2,5}}-\frac{\mathrm{d}X_{2,5}\wedge \mathrm{d}X_{3,5}}{X_{2,5} X_{3,5}}\,,
	\end{align}
	which agrees with the known result.
\end{enumerate}
Having computed $\smash{\omega_5^{\text{ABHY}}}$, we can now extract five-point amplitude $m_5$ by pulling \eqref{eq:ABHY-n=5-form} back to the appropriate subspace. Alternatively, we can use \eqref{eq:ABHY-phi-cubed-amplitude}:
\begin{align}\label{eq:ABHY-5-point-amplitude}
	m_5=\sum_{\bm{\xi}\in \mathcal{V}(I_5)} A'_5(\bm{\xi})\,,
\end{align}
where
\begin{align}
	 A'_5(\bm{z})= \frac{(1-z_2)z_3}{z_2 (z_2-z_3)(z_3-1)} \left|\frac{\partial\bm{f}}{\partial\bm{z}}\right|^{-1}.
\end{align}
We will again calculate $\mathcal{I}_{5\ast} A'_5$ using the duality of global residues which means that we can reuse the dual basis in \eqref{eq:ABHY-n=5-dual-basis} and the polynomial inverses in \eqref{eq:ABHY-n=5-poly-inverses}. To evaluate \eqref{eq:ABHY-5-point-amplitude} using \eqref{eq:gr-final-r} we first compute
\begin{align}
&\left(\overbar{(1-z_2)z_3(z_2)_\text{inv}(z_2-z_3)_\text{inv}(z_3-1)_\text{inv}}^\mathcal{G}\right)_1 =\\
&\Delta_1\left[\frac{1}{X_{1,3} X_{1,4}}+ \frac{1}{X_{1,4} X_{2,4}}+\frac{1}{X_{2,4} X_{2,5}}+\frac{1}{X_{1,3} X_{3,5}}+\frac{1}{X_{2,5} X_{3,5}}\right].\nonumber
\end{align}
which when multiplied by $\mu_1={1}/{\Delta_1}$ gives the correct five-point amplitude
\begin{align}
m_5=\sum_{\bm{\xi}\in\mathcal{V}(\mathcal{I}_5)}A_5(\bm{\xi}) = \frac{1}{X_{1,3} X_{1,4}}+ \frac{1}{X_{1,4} X_{2,4}}+\frac{1}{X_{2,4} X_{2,5}}+\frac{1}{X_{1,3} X_{3,5}}+\frac{1}{X_{2,5} X_{3,5}}\,.
\end{align}

\subsection{Momentum Amplituhedron meets ABHY Associahedron}\label{sec:examples-MOM-ABHY}
As the second application of our methods, in this section we consider examples of pushforwards of rational differential forms which are not top-dimensional. Our examples are taken from \cite{Damgaard:2020eox}; this paper explores a remarkable connection between the canonical forms of the ABHY associahedron and the momentum amplituhedron. In these examples, there are a number of different spaces and maps involved which we will define below.

The momentum amplituhedron $\mathcal{M}_{n,k}$, which is defined for $2\le k \le n-2\ge 2$, is a positive geometry whose canonical form $\Omega_{n,k}\coloneqq\Omega(\mathcal{M}_{n,k})$ encodes the $n$-particle N\textsuperscript{$k-2$}MHV superamplitude for so-called $\mathcal{N}=4$ supersymmetric Yang-Mills (SYM) theory, expressed in terms of spinor-helicity variables \cite{Damgaard:2019ztj}. The momentum amplituhedron form $\Omega_{n,k}$ is a top-form on $\mathcal{M}_{n,k}$ where the latter is a particular $(2n-4)$-dimensional region inside \emph{on-shell space $\mathbb{O}_n$}. On-shell space is the space of spinor-helicity variables. It is parametrized by two real $2\times n$ matrices $(\lambda,\tilde\lambda)$, each of which is understood modulo $\mathsf{SL}(2;\mathbb{R})$, and together they satisfy momentum conservation $\lambda\cdot\tilde\lambda^T = 0_{2\times2}$. Therefore $\dim(\mathbb{O}_n) = 4n -10$.

There is a well-known scaling of spinor-helicity variables induced by the following \emph{torus action} of the \emph{little group}
\begin{align}\label{eq:MOM-ABHY-torus-action}
	\lambda_i\mapsto t_i \lambda_i\,,\qquad\tilde\lambda_i\mapsto t_i^{-1}\tilde\lambda_i\,,
\end{align}
where $\bm{t}=(t_1,\ldots,t_n)\in\mathbb{R}_{>0}^n$. On-shell space $\mathbb{O}_n$ modulo this torus action defines a $(3n-10)$-dimensional space called the \emph{little group (LG) invariant space $\mathbb{L}_n$}. One of the simplest ways to parametrize $\mathbb{L}_n$ is to first focus on $\lambda$ (a real $2\times n$ matrix), to use the $\mathsf{SL}(2)$ symmetry to fix $3$ of its degrees of freedom, and to use the torus action to rescale each of its columns independently so that each column contains a $1$. Consequently $\lambda$ is parametrized by $n-3$ variables and gives a chart for the real-slice of the moduli space of the $n$-punctured Riemann sphere. One explicit realization of this parametrization was eluded to in the previous section, but in this section we will instead use the realization given by \emph{Fock--Goncharov (FG)} \cite{fock2006moduli}:
\begin{align}\label{eq:MOM-ABHY-lambda-FG}
	\lambda = \begin{pmatrix}
0 & 1 & 1 & 1 & 1& \cdots & 1\\
-1 & 0 & 1 & 1+z_1 & 1+ z_1+z_1z_2 &\cdots & 1 + z_1 + \ldots + z_1z_2\cdots z_{n-3}
	\end{pmatrix},
\end{align}
where $\lambda\in\mathsf{Mat}_{2,n}(\mathbb{Q}[z_1,\ldots,z_{n-3}])$. We can then fix $\tilde\lambda$ (another real $2\times n$ matrix) by using the $\mathsf{SL}(2)$ symmetry to fix $3$ degrees of freedom and to remove another $4$ degrees of freedom by imposing momentum conservation. Consequently $\tilde\lambda$ is parametrized by $2n-7$ variables, call them $z_{n-2},\ldots,z_{3n-10}$ (in addition to the $n-3$ FG variables). Moreover, it is always possible to choose the $\mathsf{SL}(2)$ transformation such that $\tilde\lambda\in\mathsf{Mat}_{2,n}(\mathbb{Q}[z_1,\ldots,z_{3n-10}])$. We will refer to this type of parametrization for $\mathbb{L}_n$ as the \emph{extended Fock--Goncharov (eFG)} parametrization.

In \cite{Damgaard:2020eox}, the authors defined the \emph{reduced momentum amplituhedron form} $\omega_{n,k}$. It is obtained from the momentum amplituhedron form $\Omega_{n,k}$ via the pullback  given by the torus action:
\begin{align}\label{eq:MOM-ABHY-omega}
	\left.\Omega_{n,k}\right|_{\eqref{eq:MOM-ABHY-torus-action}} = \Omega(\mathbb{RP}^n)\wedge \omega_{n,k}+\mathcal{O}(d^{n-2}\bm{t})\,,
\end{align}
where
\begin{align}\label{eq:MOM-ABHY-Omega-RP}
	\Omega(\mathbb{RP}^n) = \sum_{i=1}^n(-1)^{n-i}\mathrm{d}\log t_1\wedge\ldots\wedge\mathrm{d}\log t_{i-1}\wedge\mathrm{d}\log t_{i+1}\wedge\ldots\wedge\mathrm{d}\log t_n\,,
\end{align}
is the canonical form on real projective space $\mathbb{RP}^n$ and $\mathcal{O}(d^{n-2}\bm{t})$ denotes terms with sub-leading degree in $d\bm{t}$. By definition, $\omega_{n,k}$ is defined on $\mathbb{L}_n$ and is an $(n-3)$-form.

There is a natural set of equations, defined in \cite{Damgaard:2020eox}, which relates $\mathbb{L}_n$ to a particular subvariety of the kinematic space of $n$-particle Mandelstam invariants called the \emph{Gram-determinant subvariety $\mathbb{G}_n$}. Its definition is not important to our discussion, but can be found in \cite{Damgaard:2020eox}. Importantly, any chart for $\mathbb{G}_n$ can be expressed in terms of any $(3n-10)$-element subset of $n$-particle planar Mandelstam variables. Let
\begin{align}\label{eq:MOM-ABHY-X-G}
	\tilde{\mathcal{X}}_n \coloneqq \{\tilde{X}_{i,j}:(i,j)\in G_n\}\,,
\end{align}
for some choice of indexing set $G_n$ containing $3n-10$ non-adjacent pairs $(i,j)$ where $1\le i<j\le n$. We use tildes to distinguish the variables parametrizing $\mathbb{G}_n$ from those parametrizing $\mathbb{K}_n$ for convenience. Then the ideal
\begin{align}\label{eq:MOM-ABHY-ideal-L-to-G}
	\mathcal{I}^{\mathbb{L}\to\mathbb{G}}_n\coloneqq\Big\langle \tilde{X}_{i,j} - \sum_{i\le i'<j'<j}\langle i'j'\rangle[i'j']:(i,j)\in G_n\Big\rangle\subseteq\mathbb{Q}(	\tilde{\mathcal{X}}_n)[z_1,\ldots,z_{3n-10}]\,,
\end{align}
where the spinor-helicity brackets are evaluated on the eFG parametrization of $\mathbb{L}_n$, defines a map from a chart of $\mathbb{L}_n$ to a chart of $\mathbb{G}_n$. This ideal is generically zero-dimensional as we will see in examples.

Coming from the other side, there is also a natural set of equations which relates the kinematic space of $n$-particle Mandelstam invariants $\mathbb{K}_n$ to this Gram-determinant subvariety $\mathbb{G}_n$. They are given by the so-called \emph{Gram-determinant conditions}. Again, the details are not important to our discussion, but can be found in  \cite{Damgaard:2020eox}. These Gram-determinant conditions define another generically zero-dimensional ideal which we denote by $\mathcal{I}^{\mathbb{K}\to\mathbb{G}}_n\subseteq\mathbb{Q}(\tilde{\mathcal{X}}_n)[\mathcal{X}_n]$.

In \cite{Damgaard:2020eox}, the authors showed that
\begin{align}\label{eq:MOM-ABHY-result}
	\frac{1}{\left|\mathcal{V}^{\mathbb{L}\to \mathbb{G}}_n\right|}\mathcal{I}^{\mathbb{L}\to \mathbb{G}}_{n\ast}\sum_{k=2}^{n-2}\omega_{n,k} =  	\frac{1}{\left|\mathcal{V}^{\mathbb{K}\to \mathbb{G}}_n\right|}\mathcal{I}^{\mathbb{K}\to \mathbb{G}}_{n\ast}\omega_{n}^\text{ABHY}\,.
\end{align}
where $\mathcal{V}^{\mathbb{L}\to \mathbb{G}}_n$ and $\mathcal{V}^{\mathbb{K}\to \mathbb{G}}_n$ denote the complex affine varieties associated with $\mathcal{I}^{\mathbb{L}\to \mathbb{G}}_n$ and $\mathcal{I}^{\mathbb{K}\to \mathbb{G}}_n$, respectively. We will now proceed to verify this result in examples using some of the methods developed in this paper.

\paragraph{Four-points.}

Our first and simplest example occurs when $n=4$.
Since the dimensions of $\mathbb{L}_4$ and $\mathbb{G}_4$ are the same,
\smash{$\mathcal{X}_4=\{X_{1,3},X_{2,4}\}$} and \smash{$\tilde{\mathcal{X}}_4=\{\tilde{X}_{1,3},\tilde{X}_{2,4}\}$}. Consequently \smash{$\mathcal{I}^{\mathbb{K}\to\mathbb{G}}_4 = \langle X_{1,3}-\tilde{X}_{1,3}, X_{2,4} - \tilde{X}_{2,4}\rangle$} \cite{Damgaard:2020eox} and its variety is trivially \smash{$\mathcal{V}^{\mathbb{K}\to\mathbb{G}}_4 = \{(\tilde{X}_{1,3}, \tilde{X}_{2,4})\}$}. Using the following eFG parametrization for $\mathbb{L}_4$
\begin{align}
	\lambda = \begin{pmatrix}
		0 & 1 & 1 & 1\\
		-1& 0 & 1 & 1+z_1
	\end{pmatrix},\qquad\tilde\lambda=
\begin{pmatrix}
z_2 & -z_2 & z_2 & 0 \\
z_1+1 & -1 & 0 & 1 \\
\end{pmatrix},
\end{align}
we have that \smash{$\mathcal{I}^{\mathbb{L}\to\mathbb{G}}_4=\langle\tilde{X}_{1,3} - z_1 z_2, \tilde{X}_{2,4} - z_2\rangle$} and its associated variety is easily shown to be \smash{$\mathcal{V}^{\mathbb{L}\to\mathbb{G}}_4=\{(\frac{\tilde{X}_{1,3}}{\tilde{X}_{2,4}}, \tilde{X}_{2,4})\}$}.

For $n=4$, the only valid value for $k$ is $k=2$, and the reduced momentum amplituhedron form written in terms of the above eFG-variables is given by \cite{Damgaard:2020eox}
\begin{align}
	\omega_{4,2}=\mathrm{d}\log \frac{\langle12\rangle\langle34\rangle}{\langle14\rangle\langle23\rangle}=\mathrm{d}\log z_1\,,
\end{align}
while the ABHY canonical form for $n=4$ is given by \eqref{eq:ABHY-n=4-form}
\begin{align*}
	\omega_4^\text{ABHY} = \mathrm{d}\log\frac{X_{1,3}}{X_{2,4}}\,.
\end{align*}
Since the varieties of interest both contain a single point, it is easy to evaluate the relevant pushforwards via direct substitution to confirm  \eqref{eq:MOM-ABHY-result}:
\begin{align}
\mathcal{I}^{\mathbb{L}\to\mathbb{G}}_{4\ast}\omega_{4,2} = \omega_{4}^\text{ABHY}\Big|_{X\to\tilde{X}} = \mathcal{I}^{\mathbb{K}\to\mathbb{G}}_{4\ast}\omega_{4}^\text{ABHY}\,.
\end{align}

\paragraph{Five-points.} For $n=5$ we have the following two reduced momentum amplituhedron forms
\begin{align}
	\omega_{5,2}=\mathrm{d}\log \frac{\langle12\rangle\langle34\rangle}{\langle14\rangle\langle23\rangle}\wedge\mathrm{d}\log \frac{\langle13\rangle\langle45\rangle}{\langle15\rangle\langle34\rangle}\,,\qquad
	\omega_{5,3}=\mathrm{d}\log \frac{[12][34]}{[14][23]}\wedge\mathrm{d}\log \frac{[13][45]}{[15][34]}\,,
\end{align}
defined on $\mathbb{L}_5$ and the following ABHY associahedron form \eqref{eq:ABHY-n=5-form}
\begin{align*}
	\omega_5^\text{ABHY} =
	\frac{\mathrm{d}X_{1,3}\wedge\mathrm{d}X_{1,4}}{X_{1,3}X_{1,4}}+\frac{\mathrm{d}X_{1,4}\wedge\mathrm{d}X_{2,4}}{X_{1,4}X_{2,4}}+\frac{\mathrm{d}X_{2,4}\wedge\mathrm{d}X_{2,5}}{X_{2,4}X_{2,5}}+\frac{\mathrm{d}X_{2,5}\wedge\mathrm{d}X_{3,5}}{X_{2,5}X_{3,5}}-\frac{\mathrm{d}X_{1,3}\wedge\mathrm{d}X_{3,5}}{X_{1,3}X_{3,5}}\,,
\end{align*}
defined on $\mathbb{K}_5$. In order to push forward the latter onto $\mathbb{G}_5$, we first note that $\mathcal{X}_5$ and $\tilde{\mathcal{X}}_5$ have the same size since $3\times5-10 = 5 = \frac{5(5-3)}{2}$:
\begin{align}
\mathcal{X}_5=\{X_{1,3},X_{1,4},X_{2,4},X_{2,5},X_{3,5}\}\,,\, \tilde{\mathcal{X}}_5=\{\tilde{X}_{1,3},\tilde{X}_{1,4},\tilde{X}_{2,4},\tilde{X}_{2,5},\tilde{X}_{3,5}\}\eqqcolon\{a_i\}_{i=1}^5\,.
\end{align}
In this case, the ideal of maps from $\mathbb{K}_5$ to $\mathbb{G}_5$ is given by \cite{Damgaard:2020eox}
\begin{align}
	\mathcal{I}_{5}^{\mathbb{K}\to\mathbb{G}}=\left\langle {X}_{1,3}-\tilde{X}_{1,3},{X}_{1,4}-\tilde{X}_{1,4},{X}_{2,4}-\tilde{X}_{2,4},{X}_{2,5}-\tilde{X}_{2,5},{X}_{3,5}-\tilde{X}_{3,5}\right\rangle,
\end{align}
and its variety is simply
\begin{align}
	\mathcal{V}_{5}^{\mathbb{K}\to\mathbb{G}}=\left\{\left(\tilde{X}_{1,3},\tilde{X}_{1,4},\tilde{X}_{2,4},\tilde{X}_{2,5},\tilde{X}_{3,5}\right)\right\}.
\end{align}
Consequently
\begin{align}
\mathcal{I}_{5\ast}^{\mathbb{K}\to\mathbb{G}}\omega_5^\text{ABHY} = \omega_5^\text{ABHY} \Big|_{X\to\tilde{X}}\,.
\end{align}
On the other hand, if we parametrize $\mathbb{L}_5$ with the following eFG variables
\begin{align}
	\lambda = \begin{pmatrix}
			0 & 1 & 1 & 1 & 1 \\
			-1 & 0 & 1 & 1+z_1 & 1+z_1(1+z_2) \\
	\end{pmatrix},\,\tilde\lambda=
\begin{pmatrix}
z_3+(1+z_1) z_4 & -z_3-z_4 & z_3 & z_4 & 0 \\
1+z_1 (1+z_2)+z_5 & -1-z_5 & z_5 & 0 & 1 \\
\end{pmatrix},
\end{align}
then the ideal of maps from $\mathbb{L}_5$ to $\mathbb{G}_5$ is given by
\begin{align}
	\begin{split}
\mathcal{I}_{5}^{\mathbb{L}\to\mathbb{G}}=
\Big\{&\tilde{X}_{1,3}-z_1 ((z_2+1) z_3+(z_2-z_5)z_4),\\
&\tilde{X}_{1,4}-z_1 z_2 z_4,\tilde{X}_{2,4}-z_3+z_4 z_5,\tilde{X}_{2,5}-z_3-(z_1+1) z_4,\tilde{X}_{3,5}+z_1 z_4 z_5\Big\}\,.
	\end{split}
\end{align}
Its corresponding set of common zeros contains the following two algebraic solutions involving square roots:
\begin{align}
	&\mathcal{V}^{\mathbb{L}\to\mathbb{G}}_5 =\\
	\nonumber &\left\{
	\begin{array}{l}
	z_1=\frac{\pm\sqrt{\Delta}-\tilde{X}_{1,3} \left(\tilde{X}_{2,4}-\tilde{X}_{2,5}\right)-\tilde{X}_{1,4} \left(\tilde{X}_{2,5}-\tilde{X}_{3,5}\right)-\tilde{X}_{2,4} \tilde{X}_{3,5}}{2 \tilde{X}_{2,4} \left(\tilde{X}_{1,4}-\tilde{X}_{2,4}+\tilde{X}_{2,5}\right)},\\
	z_2=\frac{\pm\sqrt{\Delta}+\tilde{X}_{1,3} \left(\tilde{X}_{2,4}-\tilde{X}_{2,5}\right)+\tilde{X}_{1,4} \left(\tilde{X}_{2,5}+\tilde{X}_{3,5}\right)-\tilde{X}_{2,4} \tilde{X}_{3,5}}{2 \tilde{X}_{2,5} \tilde{X}_{3,5}},\\
	z_3=\frac{\mp\sqrt{\Delta}+\tilde{X}_{1,3} \left(\tilde{X}_{2,4}+\tilde{X}_{2,5}\right)-\tilde{X}_{1,4} \left(\tilde{X}_{2,5}-\tilde{X}_{3,5}\right)-\tilde{X}_{2,4} \tilde{X}_{3,5}}{2 \tilde{X}_{1,3}},\\
	z_4=\frac{\left(\tilde{X}_{1,4}-\tilde{X}_{2,4}\right) \left(\mp\sqrt{\Delta}+\tilde{X}_{1,3} \left(\tilde{X}_{2,4}+\tilde{X}_{2,5}\right)-\tilde{X}_{1,4} \left(\tilde{X}_{2,5}-\tilde{X}_{3,5}\right)-\tilde{X}_{2,4} \tilde{X}_{3,5}\right)+2\tilde{X}_{1,3}\tilde{X}_{2,4} \tilde{X}_{2,5}}{2\tilde{X}_{1,3}\left(\tilde{X}_{1,3}-\tilde{X}_{1,4}+\tilde{X}_{2,4}\right)},\\
	z_5=\frac{\mp\sqrt{\Delta}-\tilde{X}_{1,3} \left(\tilde{X}_{2,4}-\tilde{X}_{2,5}\right)-\tilde{X}_{1,4} \left(\tilde{X}_{2,5}+\tilde{X}_{3,5}\right)+\tilde{X}_{2,4} \tilde{X}_{3,5}}{2 \tilde{X}_{1,4} \tilde{X}_{2,5}}
	\end{array}\right\},
\end{align}
where
\begin{align}
	\Delta =& (\tilde{X}_{1,3} \tilde{X}_{2,4}-\tilde{X}_{1,3} \tilde{X}_{2,5}+\tilde{X}_{1,4} \tilde{X}_{2,5}-\tilde{X}_{1,4} \tilde{X}_{3,5}+\tilde{X}_{2,4}\tilde{X}_{3,5})^2\\
	\nonumber &+4 \tilde{X}_{1,3} \tilde{X}_{2,4}\tilde{X}_{3,5} (\tilde{X}_{1,4}-\tilde{X}_{2,4}+\tilde{X}_{2,5}).
\end{align}
The presence of the square roots in $\mathcal{V}_{5}^{\mathbb{L}\to\mathbb{G}}$ makes it extremely difficult to directly compute the pushforwards of $\omega_{5,2}$ and $\omega_{5,3}$ via $\mathcal{I}_{5}^{\mathbb{L}\to\mathbb{G}}$. Instead of performing these pushforwards directly, we will use the methods introduced in \cref{sec:alternative-companion-matrices}. For the reader's convenience, we will explicitly follow the steps outlined in \cref{sec:computation} under method \ref{par:comp-matrices-derivatives}.

\begin{enumerate}
	\item[\ref{itm:dcm-groebner}] Choosing lex ordering with $z_1\succ\ldots\succ z_5$, the Gr\"{o}bner basis is calculated to be
\begin{align}
	&\mathcal{G}\coloneqq\mathcal{G}_\text{lex}\Big(\mathcal{I}^{\mathbb{L}\to\mathbb{G}}_5\Big)=\\
	\nonumber&\left\{\begin{array}{l}
z_1+\frac{\left(1+z_5\right) \tilde{X}_{1,4} \tilde{X}_{2,5}+\tilde{X}_{1,3} \left(\tilde{X}_{2,4}-\tilde{X}_{2,5}\right)}{\tilde{X}_{2,4} \left(\tilde{X}_{1,4}-\tilde{X}_{2,4}+\tilde{X}_{2,5}\right)},\\
z_2+\frac{z_5 \tilde{X}_{1,4}}{\tilde{X}_{3,5}},\\
z_3-\frac{z_5 \tilde{X}_{1,4} \tilde{X}_{2,5}+\tilde{X}_{1,3} \tilde{X}_{2,4}+\tilde{X}_{1,4} \tilde{X}_{3,5}-\tilde{X}_{2,4} \tilde{X}_{3,5}}{\tilde{X}_{1,3}},\\
z_4-\frac{z_5 \tilde{X}_{1,4} \tilde{X}_{2,5} \left(\tilde{X}_{1,4}-\tilde{X}_{2,4}\right)+\tilde{X}_{3,5} \left(\tilde{X}_{1,4}-\tilde{X}_{2,4}\right)^2+\tilde{X}_{1,3} \tilde{X}_{2,4} \left(\tilde{X}_{1,4}-\tilde{X}_{2,4}+\tilde{X}_{2,5}\right)}{\tilde{X}_{1,3} \left(\tilde{X}_{1,3}-\tilde{X}_{1,4}+\tilde{X}_{2,4}\right)},\\
z_5^2+\frac{z_5 \left(\tilde{X}_{1,3} \left(\tilde{X}_{2,4}-\tilde{X}_{2,5}\right)-\tilde{X}_{2,4} \tilde{X}_{3,5}+\tilde{X}_{1,4} \left(\tilde{X}_{2,5}+\tilde{X}_{3,5}\right)\right)-\left(\tilde{X}_{1,3}-\tilde{X}_{1,4}+\tilde{X}_{2,4}\right) \tilde{X}_{3,5}}{\tilde{X}_{1,4} \tilde{X}_{2,5}}
	\end{array}\right\}.
\end{align}
\item[\ref{itm:dcm-standard-basis}] By inspecting the leading monomials of polynomials in the Gr\"{o}bner basis, we immediately write down the standard basis of \smash{$Q\coloneqq\mathbb{Q}(\tilde{\mathcal{X}}_5)[z_1,\ldots,z_5]/\mathcal{I}_5^{\mathbb{L}\to\mathbb{G}}$} which is given by \smash{$\mathcal{B}\coloneqq\mathcal{B}_\text{lex}(\mathcal{I}^{\mathbb{L}\to\mathbb{G}}_5) = \{z_5,1\}$}.
\item[\ref{itm:dcm-comp}] Next, we determine the companion matrices of \smash{$\mathcal{I}^{\mathbb{L}\to\mathbb{G}}_5$} to be
\begin{subequations}
\begin{align}
	T_1&=
	\begin{pmatrix}
		-\frac{(\tilde{X}_{2,4}-\tilde{X}_{1,4}) \tilde{X}_{3,5}}{\tilde{X}_{2,4} (\tilde{X}_{1,4}-\tilde{X}_{2,4}+\tilde{X}_{2,5})} & -\frac{(\tilde{X}_{1,3}-\tilde{X}_{1,4}+\tilde{X}_{2,4}) \tilde{X}_{3,5}}{\tilde{X}_{2,4} (\tilde{X}_{1,4}-\tilde{X}_{2,4}+\tilde{X}_{2,5})} \\
		-\frac{\tilde{X}_{1,4} \tilde{X}_{2,5}}{\tilde{X}_{2,4} (\tilde{X}_{1,4}-\tilde{X}_{2,4}+\tilde{X}_{2,5})} & -\frac{\tilde{X}_{1,3} \tilde{X}_{2,4}-\tilde{X}_{1,3}\tilde{X}_{2,5}+\tilde{X}_{1,4} \tilde{X}_{2,5}}{\tilde{X}_{2,4} (\tilde{X}_{1,4}-\tilde{X}_{2,4}+\tilde{X}_{2,5})} \\
	\end{pmatrix},\\
	T_2&=
	\begin{pmatrix}
		\frac{\tilde{X}_{1,3} (\tilde{X}_{2,4}-\tilde{X}_{2,5})-\tilde{X}_{2,4} \tilde{X}_{3,5}+\tilde{X}_{1,4} (\tilde{X}_{2,5}+\tilde{X}_{3,5})}{\tilde{X}_{2,5} \tilde{X}_{3,5}} & -\frac{\tilde{X}_{1,3}-\tilde{X}_{1,4}+\tilde{X}_{2,4}}{\tilde{X}_{2,5}} \\
		-\frac{\tilde{X}_{1,4}}{\tilde{X}_{3,5}} & 0 \\
	\end{pmatrix},\\
	T_3&=
	\begin{pmatrix}
		(1-\frac{\tilde{X}_{1,4}}{\tilde{X}_{1,3}}) \tilde{X}_{2,5} & \frac{(\tilde{X}_{1,3}-\tilde{X}_{1,4}+\tilde{X}_{2,4}) \tilde{X}_{3,5}}{\tilde{X}_{1,3}} \\
		\frac{\tilde{X}_{1,4} \tilde{X}_{2,5}}{\tilde{X}_{1,3}} & \tilde{X}_{2,4}+\frac{(\tilde{X}_{1,4}-\tilde{X}_{2,4}) \tilde{X}_{3,5}}{\tilde{X}_{1,3}} \\
	\end{pmatrix},\\
	T_4&=
	\begin{pmatrix}
		\frac{\tilde{X}_{1,4} \tilde{X}_{2,5}}{\tilde{X}_{1,3}} & \frac{(\tilde{X}_{1,4}-\tilde{X}_{2,4}) \tilde{X}_{3,5}}{\tilde{X}_{1,3}} \\
		\frac{\tilde{X}_{1,4} (\tilde{X}_{1,4}-\tilde{X}_{2,4}) \tilde{X}_{2,5}}{\tilde{X}_{1,3} (\tilde{X}_{1,3}-\tilde{X}_{1,4}+\tilde{X}_{2,4})} & \frac{\tilde{X}_{3,5} (\tilde{X}_{1,4}-\tilde{X}_{2,4}){}^2+\tilde{X}_{1,3} \tilde{X}_{2,4} (\tilde{X}_{1,4}-\tilde{X}_{2,4}+\tilde{X}_{2,5})}{\tilde{X}_{1,3} (\tilde{X}_{1,3}-\tilde{X}_{1,4}+\tilde{X}_{2,4})} \\
	\end{pmatrix},\\
	T_5&=
	\begin{pmatrix}
		\frac{\tilde{X}_{1,3} (\tilde{X}_{2,5}-\tilde{X}_{2,4})+\tilde{X}_{2,4} \tilde{X}_{3,5}-\tilde{X}_{1,4} (\tilde{X}_{2,5}+\tilde{X}_{3,5})}{\tilde{X}_{1,4} \tilde{X}_{2,5}} & \frac{(\tilde{X}_{1,3}-\tilde{X}_{1,4}+\tilde{X}_{2,4}) \tilde{X}_{3,5}}{\tilde{X}_{1,4} \tilde{X}_{2,5}} \\
		1 & 0 \\
	\end{pmatrix},
\end{align}
from which one can easily calculate their derivatives with respect $\tilde{\mathcal{X}}_5$, as explained in \ref{itm:dcm-comp-der}.
\end{subequations}
\end{enumerate}
For the remainder of the steps in method \ref{par:comp-matrices-derivatives}, we will specialize to $\omega_{5,2}$; the computation for $\omega_{5,3}$ yields the same result \cite{Damgaard:2020eox}. In terms of the eFG variables introduced above, $\omega_{5,2}$ is expressed as
\begin{align}
	\omega_{5,2} = \mathrm{d}\log z_1\wedge\mathrm{d}\log z_2 = \underline{\omega}(\bm{z})\mathrm{d}z_1\wedge\mathrm{d}z_2,\qquad\text{where}\qquad\underline{\omega}(\bm{z})=\frac{1}{z_1z_2}\,.
\end{align}
\begin{enumerate}
	\item[\ref{itm:dcm-rat-fn-on-cm}] Evaluating $\underline{\omega}$ on companion matrices yields
	\begin{align}
		\underline{\omega}(\bm{T})=
		\begin{pmatrix}
			\frac{\tilde{X}_{2,5}}{\tilde{X}_{1,3}} & \frac{(\tilde{X}_{1,4}-\tilde{X}_{2,4}) \tilde{X}_{3,5}}{\tilde{X}_{1,3} \tilde{X}_{1,4}} \\
			\frac{(\tilde{X}_{1,4}-\tilde{X}_{2,4}) \tilde{X}_{2,5}}{\tilde{X}_{1,3} (\tilde{X}_{1,3}-\tilde{X}_{1,4}+\tilde{X}_{2,4})} & \frac{\tilde{X}_{3,5} (\tilde{X}_{1,4}-\tilde{X}_{2,4})^2+\tilde{X}_{1,3} \tilde{X}_{2,4} (\tilde{X}_{1,4}-\tilde{X}_{2,4}+\tilde{X}_{2,5})}{\tilde{X}_{1,3} \tilde{X}_{1,4} (\tilde{X}_{1,3}-\tilde{X}_{1,4}+\tilde{X}_{2,4})} \\
		\end{pmatrix}\,.
\end{align}
\item[\ref{itm:dcm-jac-factor-on-cm-derivatives}] For each $2$-element subset $J=\{j_1,j_2\}\in\binom{[5]}{2}$ we compute
\begin{align*}
	\sum_{\sigma\in S_2}\frac{\partial T_{\sigma(1)}}{\partial a_{j_1}}\frac{\partial T_{\sigma(2)}}{\partial a_{j_2}}\,.
\end{align*}
For example, for $J=\{1,2\}$ we have
\begin{align}
\sum_{\sigma\in S_2}\frac{\partial T_{\sigma(1)}}{\partial a_{1}}\frac{\partial T_{\sigma(2)}}{\partial a_{2}} =
\begin{pmatrix}
	\frac{1}{\tilde{X}_{2,4} (\tilde{X}_{1,4}-\tilde{X}_{2,4}+\tilde{X}_{2,5})} & 0 \\
	\frac{\tilde{X}_{2,4}-\tilde{X}_{2,5}}{\tilde{X}_{2,4} (\tilde{X}_{1,4}-\tilde{X}_{2,4}+\tilde{X}_{2,5}) \tilde{X}_{3,5}} & 0 \\
\end{pmatrix}.
\end{align}
The matrix expressions for other values of $J$ are easily computed.
\item[\ref{itm:dcm-trace}] Finally, we compute the coefficients of the pushforward, denoted by \smash{$\mathcal{I}^{\mathbb{L}\to\mathbb{G}}_{5\ast}\underline{\omega}_{J}$}, via
\begin{align*}
	\text{Tr}\left[\underline{\omega}(\bm{T})\sum_{\sigma\in S_2}\frac{\partial T_{\sigma(1)}}{\partial a_{j_1}}\frac{\partial T_{\sigma(2)}}{\partial a_{j_2}}\right].
\end{align*}
For example, for $J=\{1,2\}$ we find that the coefficient of $\mathrm{d}a_1\wedge\mathrm{d}a_2 = \mathrm{d}\tilde{X}_{1,3}\wedge\mathrm{d}\tilde{X}_{1,4}$ is given by
\begin{align}
	\mathcal{I}^{\mathbb{L}\to\mathbb{G}}_{5\ast}\underline{\omega}_{\{1,2\}}= \frac{1}{\tilde{X}_{1,3}\tilde{X}_{1,4}}\,,
\end{align}
as expected. Similarly the other coefficients of the pushforward can be computed.
\end{enumerate}
Consequently, we are able to confirm that
\begin{align}
	\frac{1}{2}\mathcal{I}^{\mathbb{L}\to\mathbb{G}}_{5\ast}(\omega_{5,2}+\omega_{5,3}) =\omega_{5}^\text{ABHY}\Big|_{X\to\tilde{X}} = \mathcal{I}^{\mathbb{K}\to\mathbb{G}}_{5\ast}\omega_{5}^\text{ABHY}\,.
\end{align}

\section{Conclusions and Outlook}\label{sec:conclusions}

In this paper we have explored ways to calculate pushforwards of rational differential forms through the common zeroes of a set of equations. In particular, when these equations are polynomials we have shown that we can leverage the tools from computational algebraic geometry to calculate the pushforward without explicitly finding the set of common zeroes of these polynomials. To this end, we discussed three specific methods to do these calculations. We expect that these methods will provide us with both conceptual and practical tools for the calculations of pushforwards that had been previously not possible.

As we pointed out in the main text, a natural place where the methods developed in this paper can be applied is to the study of positive geometries. As conjectured in \cite{Arkani-Hamed:2017mur,He2022}, the canonical form of the ABHY associahedron, the momentum amplituhedron, and the orthogonal momentum amplituhedron can all be found as pushforwards of (a suitable extensions of) the world-sheet Parke-Taylor form through the solutions of the scattering equations. Using the methods we developed in this paper, it should be feasible to extend the range of canonical forms for which direct calculations are possible using this process. This could further be applied to find the canonical forms of new positive geometries, such as the conjectural `momentum amplituhedron in six dimensions' \cite{He2022}.

\begin{figure}[t]
	\centering
	\includegraphics[width=\textwidth]{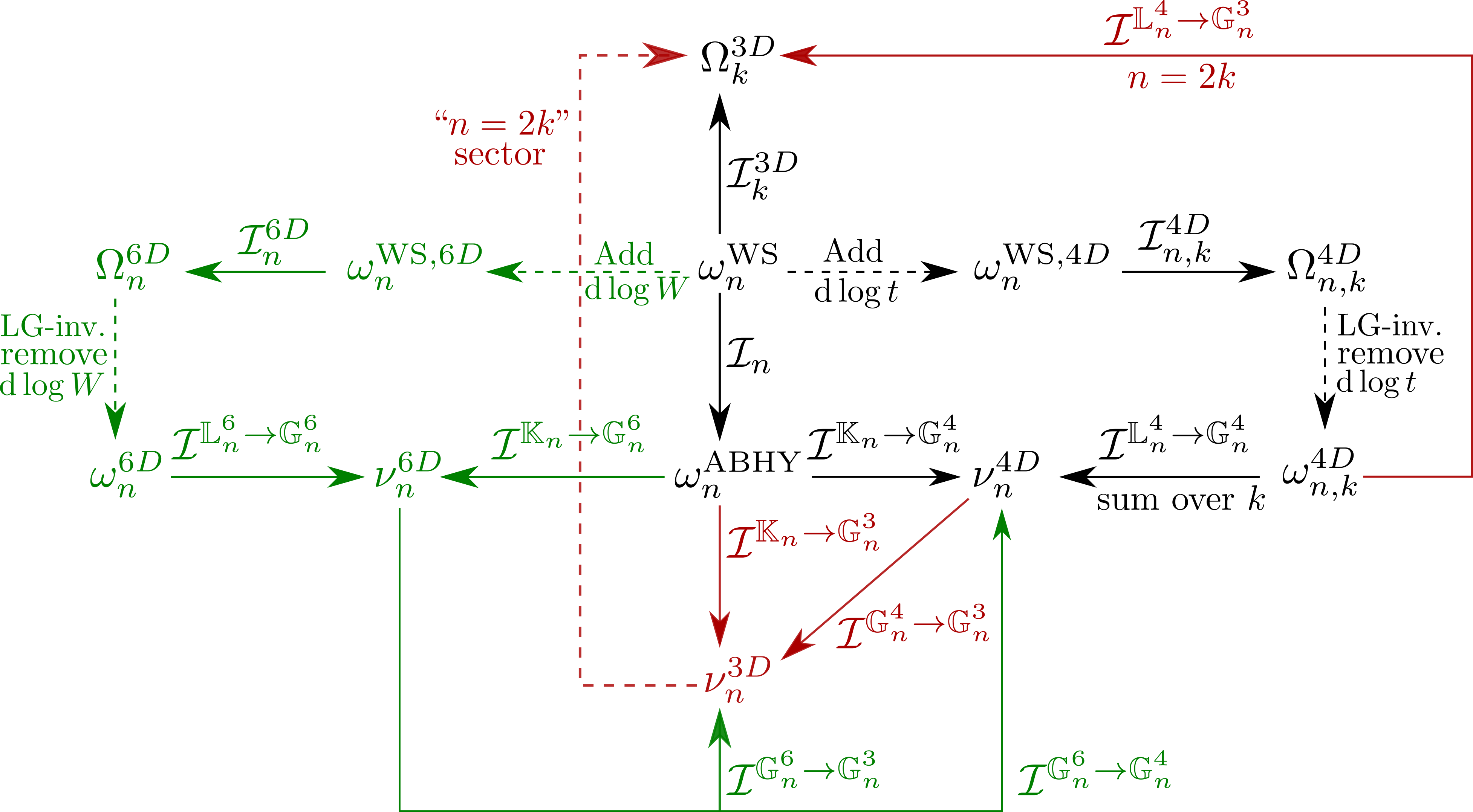}
	\caption{A web indicating the (conjectural) relations between different canonical forms. Solid lines indicate differential forms related by a pushforward. We indicated with black lines the connections that have already been investigated in the literature. Green lines depend on the hypothetical 6D momentum amplituhedron. The differential forms that appear in this web are as follows: $\omega_n^{\text{WS},4D}$ and $\omega_n^{\text{WS},6D}$ are extensions of $\omega_n^{\text{WS}}$ to include the little group in 4 and 6 dimensions, respectively. $\Omega_k^{3D},\,\Omega_{n,k}^{4D}$, and $\Omega_n^{6D}$ indicate the canonical forms of the relevant momentum amplituhedron in 3, 4, and 6 dimensions, respectively. $\omega_{n,k}^{4D}$ and $\omega_n^{6D}$ are the reduced momentum amplituhedron forms in 4D and 6D, where the little group dependence has been stripped off. $\nu_n^{q D}$ is the ABHY associahedron canonical form when restricted to $q$-dimensional kinematics. The ideals that define the pushforwards are as follows: $\mathcal{I}_{k}^{3D}, \mathcal{I}_{n,k}^{4D}$, and $\mathcal{I}_{n}^{6D}$ are the ideals generated by the $n$-particle scattering equations in 3, 4, and 6 dimensions, and $\mathcal{I}^{\mathbb{A}\to\mathbb{B}}$ is the ideal generated by the equations relating the kinematic spaces $\mathbb{A}$ and $\mathbb{B}$. $\mathbb{K}_n$ is the space of Mandelstam variables of $n$-particle scattering in arbitrary dimensions, $\mathbb{L}_n^d$ is the little group invariant space for $n$ particles in $d$ dimensions, and $\mathbb{G}_n^d$ is the Gram-determinant subvariety of $\mathbb{K}_n$ for $d$-dimensional scattering.}
	\label{fig:conclusions-form-web}
\end{figure}

Furthermore, it was observed in \cite{Damgaard:2020eox} and reviewed in \cref{sec:examples-MOM-ABHY} that the canonical forms of the momentum amplituhedron and the ABHY associahedron are closely related. A natural next step would be to provide further links of  these positive geometries to the orthogonal momentum amplituhedron and the 6D momentum amplituhedron. A conjectural web of connections, that builds on the diagram in the conclusions of \cite{He2022}, is shown in \cref{fig:conclusions-form-web}. By calculating the pushforwards indicated in this web, these statements can be verified and the precise nature of the connections between these positive geometries can be better understood.

Although the applications to positive geometries were our main motivation, the techniques developed in this paper can be easily applied to calculations of pushforwards that can be found in other corners of mathematics and physics, far beyond the scope of scattering amplitudes.

Lastly, we note that the three methods studied in this paper all require the calculation of Gr\"obner bases, which forms the main bottleneck in applying our methods to more involved examples. One possible direction would be to improve the efficiency of Gr\"obner basis calculations in this context. Alternatively, it would be very interesting to find alternative ways to calculate the pushforward (or, equivalently, summing rational functions) through the zeroes of a set of polynomials that altogether sidestep the need to calculate any Gr\"obner bases.

\section*{Acknowledgements}
The authors are grateful to Yang Zhang for providing useful comments and references.

\appendix

\section{Results from Computational Algebraic Geometry}\label{sec:theorems}
In this section we list the results from algebraic geometry which are referenced in this paper to make the paper more self-contained.

\begin{theorem}[Stickelberger's Theorem \cite{sturmfels2002solving}]\label{thm:stickelberger}
	The $d$ complex zeros of a zero-dimensional ideal $\mathcal{I}\subseteq\mathbb{C}[z_1,\ldots,z_n]$ are the vectors of simultaneous eigenvalues $\bm{\lambda}=(\lambda_1,\ldots,\lambda_n)$ of the companion matrices $T_1,\ldots,T_n$ of $\mathcal{I}$:
	\begin{align}
		\mathcal{V}(\mathcal{I})=\left\{\bm{\lambda}\in\mathbb{C}^n\,|\,\exists_{\bm{v}\in\mathbb{C}^d\setminus\{\bm{0}\}}\forall_{i\in[n]}: T_i\cdot\bm{v} = \lambda_i\,\bm{v}\right\}.
	\end{align}
\end{theorem}

\begin{theorem}[Global Duality Theorem \cite{cattani2005introduction}]\label{thm:global-duality}
	Let $\mathcal{I}=\langle f_1,\ldots,f_n\rangle\subseteq\mathbb{C}[z_1,\ldots,z_n]$ be a zero-dimensional ideal and let $Q=\mathbb{C}[z_1,\ldots,z_n]/\mathcal{I}$ be the corresponding quotient ring. Then the symmetric inner-product
	\begin{align}
		\langle\bullet,\bullet\rangle:Q\times Q\to\mathbb{C},\; (p_1,p_2)\mapsto\mathrm{Res}(p_1\,p_2)\,,
	\end{align}
	where $\mathrm{Res}$ computes the global residue with respect to the $f_1,\ldots,f_n$, is non-degenerate.
\end{theorem}

\begin{theorem}[Hilbert's Weak Nullstellensatz \cite{cox2013ideals}]\label{thm:nullstellensatz}
	Let $\mathcal{I}\subseteq \mathbb{C}[z_1,\ldots,z_n]$ be an ideal satisfying $\mathcal{V}(\mathcal{I})=\emptyset$. Then $\mathcal{I}=\mathbb{C}[z_1,\ldots,z_n]$.
\end{theorem}

\begin{theorem}[Shape Lemma \cite{sturmfels2002solving}]\label{thm:shape}
	Let $\mathcal{I}$ be a zero-dimensional radical ideal in $\mathbb{C}[z_1,\ldots,z_n]$ such that all $d$ complex roots of $\mathcal{I}$ have distinct $z_n$ coordinates. Then the unique reduced Gr\"{o}bner basis $\mathcal{G}$ of $\mathcal{I}$ with respect to the lexicographic order where $z_1\succ\ldots\succ z_n$ has the shape
	\begin{align}
		\mathcal{G} = \left\{z_1-q_1(z_n),\ldots,z_{n-1}-q_{n-1}(z_n),r(z_n)\right\},
	\end{align}
	where $r\in\mathbb{Q}[z_n]$ has degree $d$ and each $q_i\in\mathbb{Q}[z_n]$ has degree strictly less than $d$.
\end{theorem}

\begin{theorem}[Elimination Theorem \cite{cox2013ideals}]\label{thm:elimination}
	Let $\mathcal{I}$ be an ideal in $\mathbb{C}[z_1,\ldots,z_n]$ and let $\mathcal{G}$ be a Gr\"{o}bner basis of $\mathcal{I}$ with respect to the lexicographic order where $z_1\succ\ldots\succ z_n$. Then for every $0\le\ell\le n$, $\mathcal{G}\cap\mathbb{C}[z_{\ell+1},\ldots,z_n]$ is a Gr\"{o}bner basis  of the $\ell$\textsuperscript{th} elimination ideal $\mathcal{I}\cap\mathbb{C}[z_{\ell+1},\ldots,z_n]$.
\end{theorem}

\bibliographystyle{nb}

\bibliography{push-forward}

\end{document}